\documentclass[useAMS,usenatbib]{mn2e}
\usepackage{times}
\usepackage{graphicx}

\newcommand{\figce}[3]{ 
\begin{figure}
\includegraphics[angle=0,width=\columnwidth]{#1}
   \caption{ { } #2 \label{#3}}
  \end{figure}
}

\newcommand{\figcw}[3]{ 
  \begin{figure*}
    \includegraphics[angle=0,width=480pt,height=270pt]{#1}
    \caption{ { } #2 \label{#3}}
  \end{figure*}
}

\newcommand{\figcd}[5]{
  \begin{figure*}
    \includegraphics[angle=0,width=\columnwidth]{#1}
    \hspace*{5pt}
    \includegraphics[angle=0,width=\columnwidth]{#2}
    \caption{ {\sl (Left) } #3 {\sl (Right) } #4 \label{#5}}
  \end{figure*}
}

\def\apj{ApJ}
\def\mnras{MNRAS}
\def\aap{A\&A}
\def\msun{\mbox{M$_\odot$} }
\def\zsun{\mbox{Z$_\odot$}}

\title[Chemical consequences of a poorly sampled IMF]
{Chemical consequences of low star formation rates: 
stochastically sampling the IMF}

\author[L. Carigi and X. Hernandez]
{Leticia Carigi$^{1,2}$ and Xavier Hernandez$^{1}$ \\
$^1$ Instituto de Astronom\'{\i}a, Universidad Nacional Aut\'onoma de M\'exico,
 A.P. 70--264,  M\'exico 04510 D.F., Mexico\\
$^2$ Centre for Astrophysics, University of Central Lancashire,
Preston, Lancashire, PR1 2HE, United Kingdom	}
\begin{document}

\date{Accepted ... Received ... ; in original form ... }

\pagerange{\pageref{firstpage}--\pageref{lastpage}} \pubyear{2008}

\maketitle

\begin{abstract}
When estimating the abundances which result from a given star formation event,
it is customary to treat the IMF as a series of weight factors to be applied to the
stellar yields, as a function of mass, implicitly assuming one is dealing with an infinite population.
However, when the stellar population is small,
the standard procedure would imply the inclusion of fractional numbers of stars at certain masses.
We study the effects of small number statistics on the resulting abundances by performing  
an statistical sampling of the IMF to form a stellar population out of discrete numbers of stars. 
A chemical evolution code then follows the evolution of the population, and traces the resulting
abundances. The process is repeated to obtain an statistical distribution of the resulting
abundances and their evolution. We explore the manner in which different elements are affected, 
and how different abundances converge to the infinite population limit as the total mass increases. 
We include a discussion of our results in the context of dwarf spheroidal galaxies and show the recently 
reported internal dispersions in abundance ratios for dSph galaxies might be { partly} explained through
the stochastic effects introduced by a low star formation rate, which can account for dispersions of over
2 dex in [C/O], [N/O], [C/Fe], [N/Fe] and [O/Fe].

\end{abstract}


\section{Introduction}

The study and modeling of the chemical consequences and effects of star formation in galaxies
has traditionally grown out of chemical evolution models originally designed to explain the
observed abundances and abundance gradients measured for the Milky Way. It has since been 
customary to treat the distribution of stellar masses through a probability function, the IMF.
When dealing with a large galaxy, or a large star formation episode in general, it is therefore
justified to think of the IMF as a densely sampled probability function. In practice, this
translates into chemical evolution codes where the IMF appears as a series of weighting factors
to be applied to the heavily mass dependent stellar yields, when calculating the overall
enrichment produced by a population of stars. 

In going to small star formation events, individual star forming regions, or fractions of
small galaxies e.g. the stellar populations of local dSph's recently available to detailed
observation however, the fundamental assumption of chemical evolution codes, that the IMF
is a densely sampled probabilistic distribution function, breaks down. If we are dealing 
with a small star formation event, calculating the resulting chemical consequences will require
the explicit inclusion of the probabilistic nature of the IMF, in a regime where this function
is being only poorly and discretely sampled. 

The chemical consequences of this change in regime are amplified by the strong power law
character of the IMF, further compounded by the heavily mass weighted dependence of the 
stellar yields. In large stellar populations of fixed total mass, the actual number of 
stars with masses beyond 8 \msun might vary by only a fraction of a percent, due to the 
intrinsic variance associated with the probabilistic sampling of the IMF. The resulting 
intrinsic variance in the final yields will be correspondingly small, and is hence customarily
ignored. If the total mass being turned into stars is of up to a few thousand \msun however,
the intrinsic probabilistic variance of a standard IMF will lead one to expect only between 
0 and a few stars beyond 8 \msun. It is clear that the resulting intrinsic variance to be expected
in the final chemical yields resulting from such a population will be of a very large factor, 
with correlations and distributions heavily dependent on both the total mass, and the actual element
being studied. Considering typical efficiency factors of a few percent, a star formation episode 
resulting in a few thousand stars implies of order $10^5$ \msun in total gas mass involved.

The equivalent statistical variance in the light output of a stellar population has been
studied (e.g. Tonry \& Schneider 1988, Cervi\~{n}o \& Valls-Gabaud 2003). 
The most important practical application
of which has been the development of the surface density fluctuation
method of distance determination in galaxies, Tonry \& Schneider (1988). Similarly, such studies
lead to the realization that the intrinsic variances in populations
of even thousands of stars, in certain observed bands or emission lines, is substantial, and has to
be taken into account. Using results of population synthesis codes which assume the IMF has been densely
sampled, and which yield a unique answer for certain observed properties of a stellar population
of fixed mass and metallicity, can lead to the over-interpretation of observations 
(Cervi\~{n}o \& Valls-Gabaud 2003), and to the invoking of physical causes (changes in the IMF, 
differences in age or metallicity), when what one is observing is merely the result of a poorly
sampled IMF.

Recent observational determinations of abundances in dSph galaxies have often shown substantial
internal variations and fluctuations within single galaxies. For example, internal dispersions in
[$\alpha$/Fe] vs. [Fe/H] in observations of dSph's 
(Venn et al. 2004, see also figure 8 in this paper)
and in nearby dIrr galaxies (e.g. van Zee \& Haynes 2006) have been reported. These results have been 
interpreted as evidence of complex gas flows and galactic wind episodes 
in dSphs (Carigi et al. 2002, Marcolini et al. 2006, 
Ikuta, \& Arimoto 2002, Fenner et al. 2006, Lanfranchi \& Matteucci 2007) and
in dIrr (Pagel 1997, Carigi et al. 2006), although the high dark matter content of these 
systems (e.g. Gilmore et al. 2007), and the low star formation rates (e.g. Hernandez et al. 2000) might make 
large scale gas loss unfeasible. Internal variations in the IMF have also been proposed,
as has the possibility of having an IMF with a lower upper mass limit or a steeper index
e.g. Weidner \& Kroupa (2005). Other authors, (e.g., Goodwin \& Pagel 2005 and references therein)
have also suggested that in galaxies with low star formation rates, the upper mass limit for
stars formed might be fixed at a lower value than what is inferred for the Milky Way disk. 
Perhaps, the apparent lower mass limit in low star formation rate systems is merely the consequence of
a poor sampling of a standard IMF, due to the low mass of the star forming complexes in these 
systems, as suggested for example, by Massey (1998). More recently, Kroupa \& Weidner (2003) 
and Weidner \& Kroupa (2004)
have convincingly argued that an upper mass limit for star formation events exists. Here we  
adopt the above result, and show that even if an upper mass limit for stellar masses exists, in going to
low total mass star formation events, an effective reduced upper limit appears, linked to the total stellar
mass, as a natural result of the stochastic sampling of the IMF.

In this paper we explore the option for observed variations in abundances 
being partly the result of intrinsic statistical
variances in a discretely and poorly sampled standard IMF. It is to be expected that the stochastic
effects on the metallicities resulting from small star formation events will decrease as the initial
gas metallicity increases, and the elements produced become progressively a smaller addition to
a growing pool of already present heavy elements. Conversely, in going to lower initial
metallicities, the stochastic effects will increase, and will be felt for even larger star formation
events, becoming maximal in the cases of small star formation events occurring in gas of primordial or
very low initial metallicities, interestingly, the scenario envisioned for the first phases of
star formation in dwarf galaxies.

\figcd{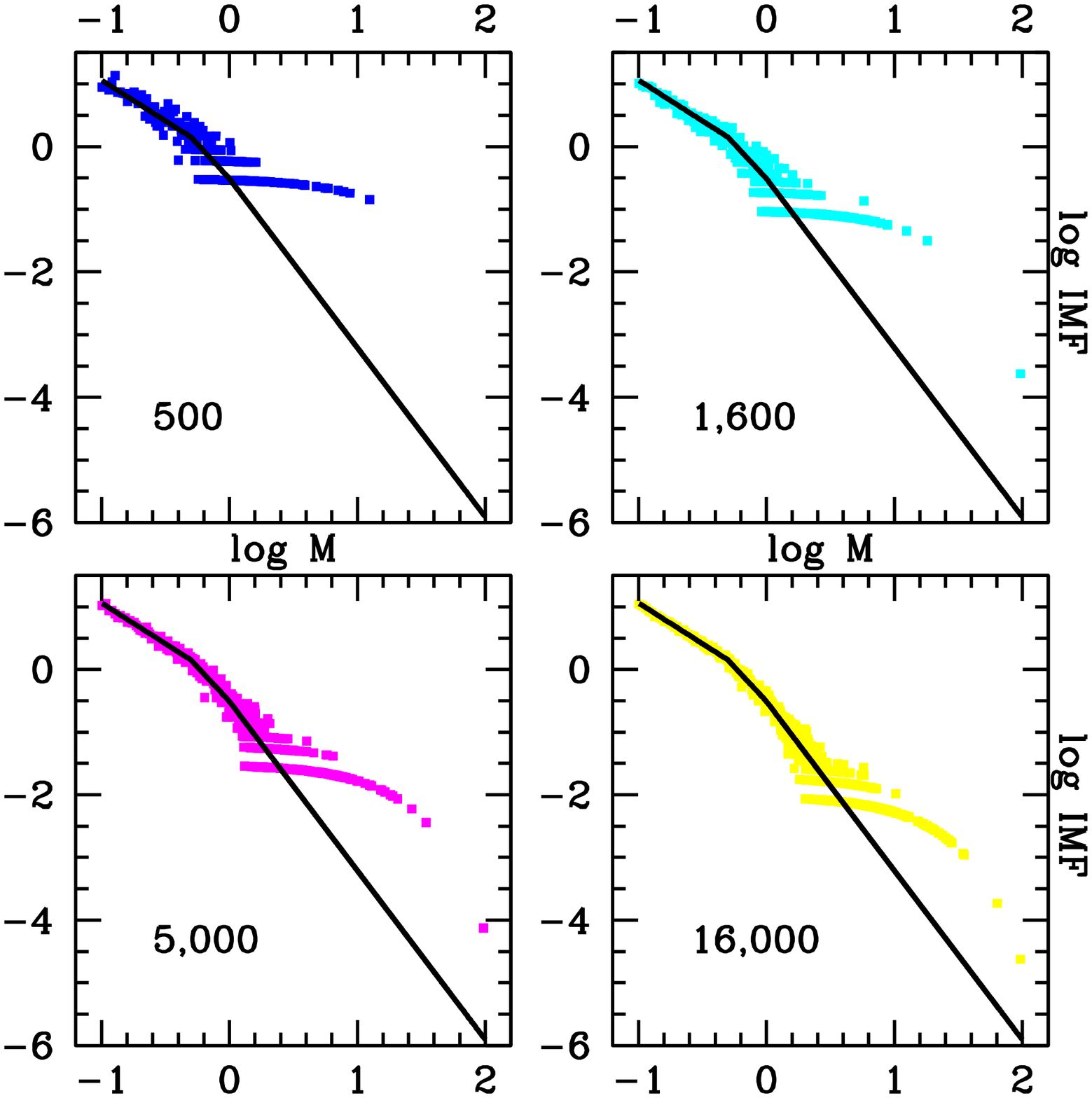}{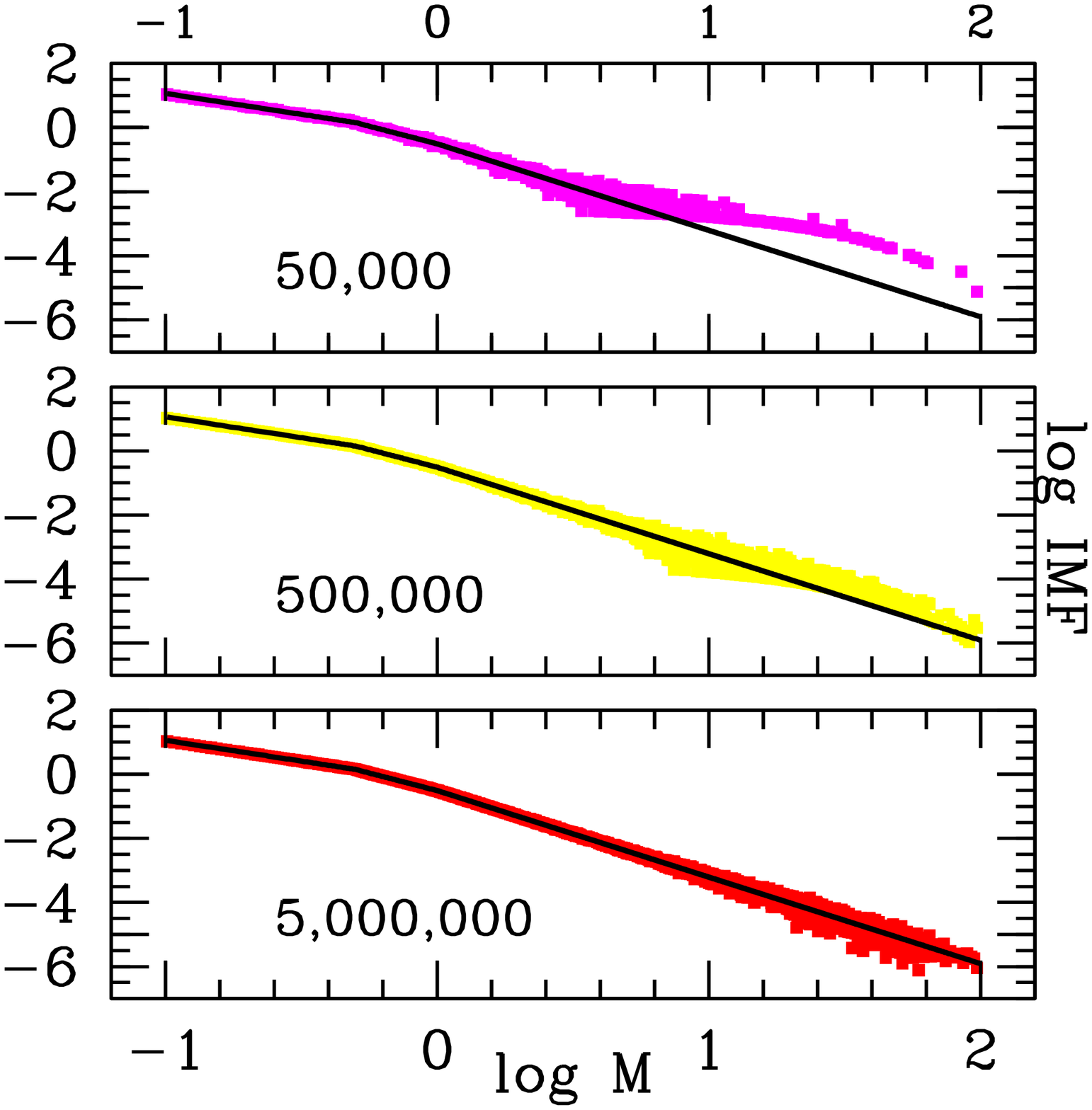}
{Particular random realizations of discrete initial mass functions for low and medium stellar total masses.
For a star formation efficiency of 5\%, the gas available and used to form stars is
$M_{stars}=0.05 \times (10^{4.0}$, $10^{4.5}$, $10^{5.0}$ and $10^{5.5}$) \msun.
The total stellar mass formed is indicated in each panel, in units of \msun.}
{Particular random realizations of discrete initial mass functions for high stellar total masses.
For a star formation efficiency of 5\%, the gas available and used to form stars is
$M_{stars}=0.05 \times (10^{6.0}$, $10^{7.0}$ and $10^{8.0}$) \msun.
The total stellar mass formed is indicated in each panel, in units of \msun.}{fig01}

Koeppen et al. (2007) calculated the stochastic effects on the resulting chemical output, of 
linking variations in the upper stellar mass limit
of a stellar population to the total stellar mass, to successfully reproduce the galactic
mass metallicity relation, in the absence of galactic winds. Here we extend the exploration of the
stochastic effects of the IMF sampling, beyond the consideration of continuous IMF with changing
upper mass limits, to fully stochastic collections of stars, where the probabilistic effects of the
sampling appear at all masses. Although the strongest effect on the resulting metallicities comes from
the changes in the effective upper mass, the spread in resulting metallicities for populations
of a fixed mass, appears only in this completely probabilistic approach, across the entire
stellar mass range being considered. In the case of the chemical consequences of these effects, Cervi\~no \& Molla (2002) and 
Cervi\~no \& Luridiana (2006) have developed
a thorough theoretical probabilistic formalism, and show the effects to be substantial, even for
stellar populations where one might naively think numbers offer 'statistical samples'. In this paper
we develop an alternative approach, more statistical than probabilistic, gaining 
much in clarity and directness, albeit tending somewhat more to the descriptive than the predictive.

We simulate stellar populations having various total masses by sampling directly an assumed IMF, 
this produces a discrete collection of stars, which is then used as input for a chemical
evolution code. We explore the resulting abundances as functions of the 
initial metallicity and the total mass, by following directly the yields and enrichment 
consequences of each individual star formed. Repeating the process a large number of times 
yields a distribution of metallicities resulting from the same input parameters.
Recently Cescutti (2008) applied precisely such a formalism to study the spread in
neutron capture elements for low metallicity stars in the Solar Neighborhood. We shall see that
the interesting effects described by this author can also play a part in understanding the large internal spreads in
abundance ratios observed in local dSph galaxies.

The paper is organized as follows: The construction of a discrete IMF is described in section 2, 
with the chemical evolutionary code, and an analysis of the final results given in section 3. 
A comparison of these in the case of observed abundances for galactic dSph's is presented in section 4,
and a general discussion of the temporal progression of the chemical evolutionary code appears in section 5.
Our conclusions are summarized in section 6.

\section{Setting up a discrete IMF}

The first step is the construction of a discrete IMF, consisting of a set of individual stars.
We first choose an underlying probabilistic IMF, which will be sampled to produce the many
discrete IMFs to be used. We take the fit given by Kroupa et at. (1993) to this function,
henceforth KTG. Specifically, a three power-law approximation, given by IMF $\propto m^{-\alpha}$,
with $\alpha=-1.3$ for 0.1 - 0.5 \msun, $\alpha=-2.2$ for 0.5 - 1.0 \msun and
$\alpha=-2.7$ for 0.5 - 100 \msun.

The details of our study are only slightly dependent on this choice, all our results would be 
qualitatively equivalent if any other function where used. In any case, in the present paper we aim at
identifying and describing the chemical consequences of low star formation regimes, rather than 
at making precise predictions for any particular case.

We take the KTG IMF, with a lower and upper mass limits of 0.1 \msun and 100 \msun respectively,
and integrate it to obtain the cumulative distribution, which is then normalized to 1. A random number
generator then provides numbers between 0 and 1, which are used to select stellar masses from the
above function. The process is repeated until the total stellar mass reaches some predetermined number,
and a particular realization of a discrete IMF is completed. For each of the 8 values for the 
total stellar mass explored, $0.05 \times (10^{4.0}, 10^{4.5}, 10^{5.0}, 10^{5.5} 10^{6.0}, 
10^{7.0}, 10^{8.0})$\msun a large number of such realizations were constructed, each in turn made up of
a particular collection of stars, as the random seed for each realization was taken differently.
It is these collections of stars, treated and followed individually, which will form the inputs for the
chemical evolutionary code described in the following sections.

Once the collections of stars are ready, we first compute the actual resulting discrete IMF, by binning the
obtained stars into intervals of fixed logarithmic width in mass, and compare the results to
the underlying probabilistic IMF.
Figure 1 presents one particular realization of a discrete IMF, for each one of the masses explored,
with the numbers appearing in each panel giving the total stellar mass in each such IMF. As can be seen
in the panels corresponding to low total stellar masses, the discrete IMFs appear as collections 
of points, the solid squares shown. The multi-valued appearances are due to the width of the squares,
this discrete IMFs are in fact full of gaps, evident at the higher stellar masses in each case. The detached
series of points seen at the bottom of the lower total stellar mass IMFs are defined by particular stellar
mass ranges where only one star is found in the statistical sampling of the 
underlying KTG IMF which gave rise
to the plotted discrete IMF. Similar series mark stellar masses where exactly two, three, etc. stars appeared
in each mass bin. The curved appearance of this features being due to the fixed logarithmic interval
used in the binning along the horizontal axis. 

The solid curves give the underlying probabilistic KTG IMF used in all cases. It is illustrative to see
how the discretely sampled IMFs converge to the probabilistic KTG function as the total stellar 
mass increases. Also, it is evident that even up to the case of a total stellar
mass of 16,000 \msun, corresponding to a star formation episode involving typically around $10^{5.5}$\msun
of gas,
the discrete IMF has still not converged to the underlying KTG function in the region where progenitors
of SN type II are found. This leads one to expect substantial intrinsic variance in the resulting
yields and metallicities, even at this relatively high masses. We see clearly how the low mass weighted
underlying IMF, compounded with the strong leverage afforded by the heavily high mass weighted yields,
has large potential to lead to significant intrinsic variances in final yields and metallicities
associated with star formation events in the range of even massive HII regions, and certainly small sections
of local dSph galaxies. At the low mass cases one requires to run the IMF sampling hundreds of times 
before a single, say, 50 \msun star appears in an IMF realization. Thus, in practice, one is dealing
with an effective IMF showing a low upper mass cut-off. Studies indicating low mass limits to
empirically determined IMFs, could equally well be interpreted as measures of the total mass of
the star-forming complexes one is observing, e.g. the results of Goodwin \& Pagel (2005).
We also note the results of Weidner \& Kroupa (2004) who show that for the KTG upper mass slope
in the IMF, an upper limit stellar mass necessarily exists, and explore the effects of the
maximum stellar mass appearing in a certain stellar cluster being a function of the cluster mass.
The above results are consistent with figure 1, the stochastic sampling effects giving effective
maximum stellar masses which decrease with the total stellar mass of a population, providing a 
causal mechanism for the linking of the upper stellar mass to that of the stellar population found in
Weidner \& Kroupa (2004).

Being the elements produced by Type Ia super novas (henceforth SNIa) some of the most well 
studied elements in chemical evolution studies, and being also easily measured in stars, they clearly 
are a feature which 
we are interested in deriving results for, particularly Fe. We shall assume the single degenerate 
scenario within red-giants plus white-dwarf systems for the formation of SNIa, thought to occur
when a white dwarf star accretes material from a red giant companion through Roche lobe overflow.
In that scenario, the progenitors of SNIa are binary
systems formed by two low or intermediary mass stars,
($m < 8$ \msun) and therefore the binary systems of interest will be modeled as having total 
masses in the range 2.65\msun $<m<$ 16\msun.
The lower limit of 2.65\msun comes from adding a minimum stellar mass required for the white dwarf of
0.80 \msun, and a minimum helium nucleus mass for the red giant of 1.85 \msun, see Matteucci (2001).
The SN explosion occurs at the end of the lifetime of the secondary star, $m2$ (red giant).

Following Greggio \& Renzini (1983), before entering the chemical evolution code we 
construct a set of binaries having a total combined mass given by 5 \% of the stellar mass 
between 2.7 and 16 \msun. For each binary system of total mass $mb$, which is taken to follow the 
same underlying IMF as the single stars, the secondary mass is obtained
by sampling the normalized distribution function $f(m2)=24(m2/mb)^2$, which yields
the mass $m2$ and therefore the time of the explosion for each SNIa. We hence only construct and
follow binaries in the mass range relevant for SNIa formation. Greggio (2005) presents a slightly 
revised treatment for SNIa, but comments that the changes with respect to Greggio \& Renzini (1983) 
are only minor, for which reason, we keep the prescription of the latter. Those binaries with a 
total mass between 2.65 and 16 \msun form a discrete collection of systems which will be the progenitors of SNIa.


\section{Chemical Evolution Models}\label{sec:models}

Once a discrete IMF has been constructed, as detailed in the preceding section,
the resulting set of individual stars and binaries is used as input for a full
chemical evolution model. A detailed description of the code can be found in Carigi (1994) and
Carigi et al. (2005), here we only outline the physical assumptions.

\subsection{Model Parameters}

 With the aim of
obtaining results which are easy to compare against well known chemical models, we chose to run only
closed box single burst models. 
 Closed box chemical evolution models are a benchmark against which to compare,
are easy to understand, and will serve as a simple and well known test case.
More realistic chemical evolution models might allow a fuller description and might
yield a better comparison with observations (e.g. the self consistent chemical and physical
models of Carigi et al. 2002, which include gas infall and
galactic winds), but the details and many parameters involved would make it difficult to
see clearly the consequences of the effect we are exploring, which can be clearly
grasped if the problem is simplified to isolate the effect in question. The closed-box
models we run are in effect a controlled experiment for the chemical consequences of
discrete sampling in the IMF, at the low mass regime. For example, Koppen et al. (2007)
 explore the chemical consequences of linking an upper stellar
mass limit to the total mass of a star formation event, comparing also against the case of
closed-box models, to cite one recent example of such a usage of closed box models.

The star formation of this single burst of duration
{$\Delta t= 3 \times 10^{-2}$ Gyr} is hence $SFR(t) = 0.05 M_{tot}/\Delta t$ for $t<\Delta t$
and $SFR(t) = 0.0 $ for the rest of the evolution, where the values for the total 
initial gas masses, $M_{tot}$, into which the yields
of the diverse collections of stars are being injected are:
$(10^{4.0}, 10^{4.5}, 10^{5.0}, 10^{5.5}, 10^{6.0}, 10^{7.0}$ and $10^{8.0})$\msun. 
Assuming a fixed
star formation efficiency of 0.05, this results in the total stellar masses of the discrete IMFs
of figure 1. 
{ 
The code follows the lifetime of each individual star and binary, for a fixed
total time span of 13 Gyr.
In our code, we assume the lifetimes change with the initial $Z$
using the lifetime data for $Z=$ $10^{-5}$, 0.004, and \zsun
by Meynet \& Maeder (2002), by Meynet \& Maeder (2001), and
by Meynet \& Maeder (2000), respectively.
A 0.03 Gyr lifetime corresponds to a 8-9 \msun star, for
the three metallicities of stellar population assumed in this paper,
but we do not consider enrichment by massive stars during the stellar burst.
}
We are thinking of the star formation of a dSph galaxy as the sum of many small
total mass star formation events, the mass of which can then be estimated, through requiring that resulting
internal spreads in abundances match observations.
{
We do not consider exchange of any type of material among stellar populations,
therefore we follow the evolution of the ISM for stellar populations formed
in a closed box scenario.
Similarly Cescutti (2008) consider random formation of new stars in
different regions where each region does not interact with the others.
}

For each value of the total mass, three initial metallicities are tested,
$Z_{i}=10^{-5}$ to represent a 'primordial' gas metallicity, to which we shall henceforth refer to as
'primordial metallicity', $Z_{i}=0.004$, and a solar value of $Z_{i}=0.02$, \zsun,
with respective initial helium values, $Y_{i}$, of 0.25, 0.258, and 0.275. 
The value of $Y_{i}=0.25$ was taken from WMAP,
$Y_{i}$=0.258 is obtained considering $\Delta Y/\Delta Z  = 1.97$,
 (Carigi \& Peimbert 2008), 
and $Y_{i}=0.275$ is the solar value determined 
by Anders \& Grevesee (1989) and the initial stellar helium fraction of the yields considered in this
work for \zsun (Maeder 1992). For all other heavy elements mentioned, the initial abundances were
scaled to solar values.

We shall refer to the fraction of material produced and ejected by a star
during its total lifetime as its yield. All yields are taken as dependent on
the metallicity of the stars, $Z$, except for the case of the SNIa's.
The detailed yields used here are those found by Carigi et al. (2005) to optimally
reproduce solar vicinity observed abundances, again, our conclusions, being mostly qualitative, 
are robust to this particular choice of yields. A brief description of the yields used follows.

For massive stars, those with $8 < m/\msun < 100$, we have used
yields from Meynet \& Maeder (2002), henceforth MM02,
for  $Z = 10^{-5}$ and $Z = 0.004$
and yields by Maeder (1992, henceforth M92) for $Z = 0.02$ (their high mass-loss rate).
{ As M92 do not give complete N yields, we took
N yields for \zsun by Hirschi et al. (2005).}
Since those yields are computed until pre-SN stage, we use
Fe yields by Woosley \& Weaver (1995, henceforth WW95) 
(their Models B, for 12 to 30 \msun; their Models C, for 35 to 40 \msun).
M92 and MM02 stellar evolution models assume stellar winds, 
but WW95 does not, therefore it is not valid
to combine M92 or MM02 yields with WW95 yields using the initial stellar mass.
Therefore, we have connected the $M_{CO}$, mass of the
carbon-oxygen cores,
from the Geneva group to $M_{CO}$ from WW95 using the prescription
given in  Portinari et al. (1998) to find the Fe yields for each initial stellar mass.
After these assumptions, Fe yields for $m > 40 $ \msun are similar to those
for $m = 40$ \msun, { that is, almost zero for low $Z$ or slightly negative for high $Z$.}

For  low and intermediate mass stars (LIMS), those with $0.8 \leq m/\msun \leq 8$,
we have used yields by Marigo, Bressan \& Chiosi (1996, 1998), and Portinari et al.  (1998)
for $Z=0.004$ and $Z=0.02$. 
{ Only for LIMS we use the same $Z=0.004$} yields for the stellar population of $Z = 10^{-5}$.
Yields for SNIa we have taken from Thielemann, Nomoto \& Hashimoto (1993),
who also show that in this case, yields are largely independent of $Z$.
For each set of yields, linear interpolations for different stellar masses
were made. For masses higher than those available we adopted the yields predicted for
the highest stellar mass  available. All models start at [Xi/Xj]=0.0 when Xi and Xj are 
both heavy elements, as we have scaled the initial abundances to solar, except for H and He.

In order to more clearly understand the [Xi/Xj] values obtained by our models,
we summarize the previous stellar yields by type of stars and initial metallicity,
listing, to first order, the principal producers of various elements.
O coming mostly from MS of any metallicity and LIMS with low $Z$, as a { minor contribution}.
C comes from LIMS and from massive stars between 8-12 \msun
of any metallicity and MS of solar metallicity. { N is produced by LIMS} and Fe from SNIa with
{ minor contributions} from SNII. Finally, He coming from MS and LIMS, of all metallicities.

{
Specifically, in massive stars:
Fe is formed in the last main burning by decay of Ni, its yield depends
on the explosion mechanism (WW95).
O is produced mainly in the pre-SN stage of stars of any $Z$,
but O yields are higher at low $Z$, when the stellar winds
are not important, while
C yields behave opposite to O yields, C yields for high $Z$
are higher than for low $Z$.
Massive stars with high $Z$ values have strong winds, lose a considerable
amount of C, leading to high C yields.
Due to this C loss, the stars keep only a small amount of C, needed to produce O,
and consequently their O yields are low.
N yields are higher for high $Z$ than low $Z$, because
N is synthesized from C and O during the CNO cycle
in the stellar interiors. If the star was born with high C and O abundance,
the N production is high (See M92, MM02, Hirschi et al 2005).

Moreover, LIMS synthesize large amounts of He, C and N.
O yields are null or negative for LIMS at high $Z$,
but according to Marigo et al. (1996, 1998), stars with $m<3$ \msun
and low $Z$ can  produce non negligible amounts of O
in the AGB and TP-AGB stages, due to the low stellar winds
(long duration of nucleosynthesis bursts).
C yields are lower at high $Z$, again because of the high mass loss rate;
stellar winds are intense in LIMS and stars have less mass to produce
heavy elements.
N yields increases with $Z$ in most LIMS, because
N production depends on the amount of C and O available to burn
hydrogen through the CNO cycle.
However, the N increase with $Z$ does not apply at all masses because
N synthesis decreases with $Z$ due to stellar winds.
If the mass loss rate is high, the duration and intensity of
the N nucleosynthesis are lower
(See Marigo et al. 1996, 1998; Portinari et al. 1998).
Very close binary systems of LIMS, that is progenitors of
SNIa, produce mainly Fe, and in small amounts O and C.
A SNIa ejects approximately 1.4 \msun of processed material,
44\% correspond to Fe, 10 \% to O, 2 \% to C, and the rest
to S and Si, mainly (Thielemann et al. 1993).
}

\subsection{Model Results}

\figce{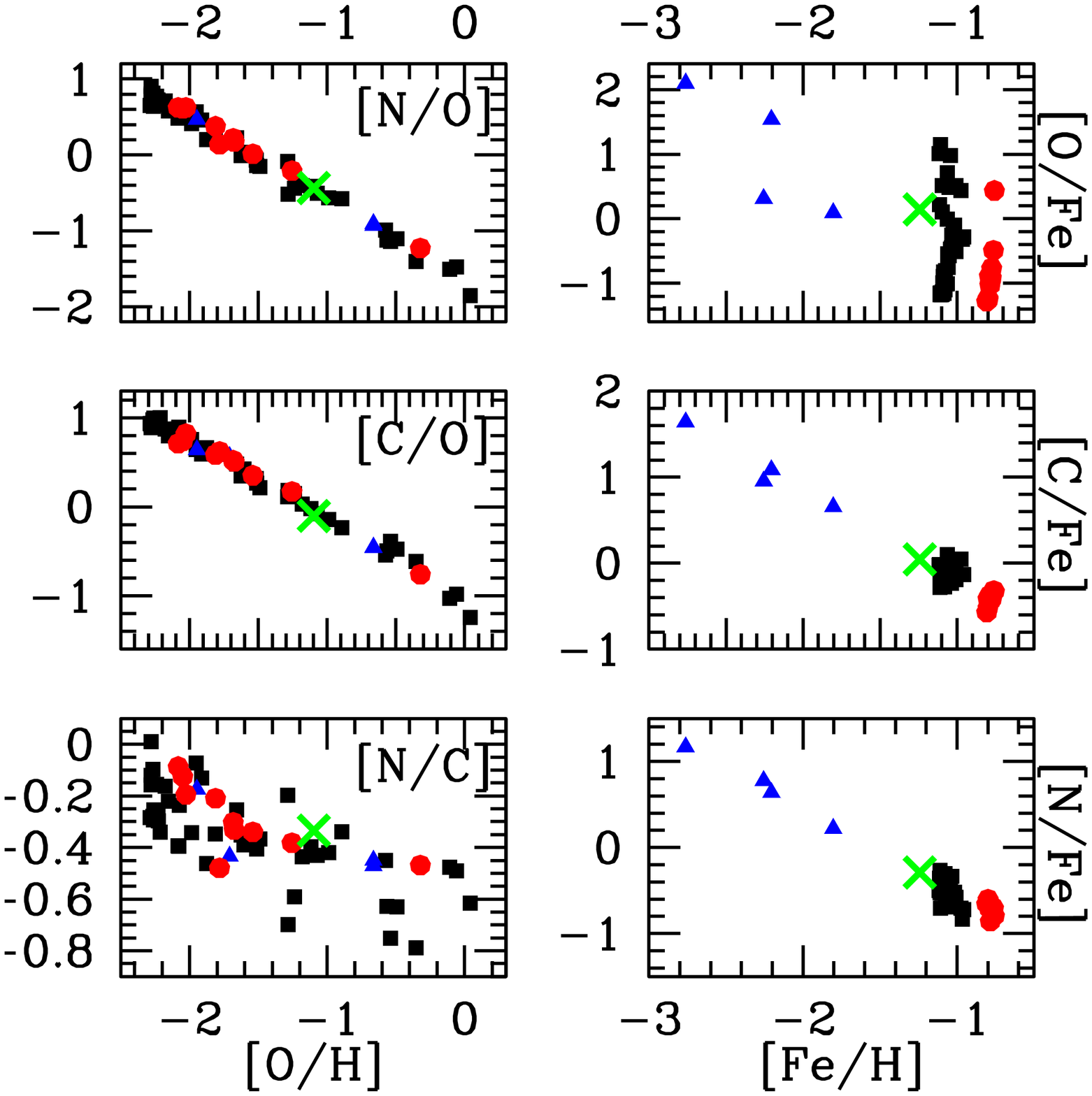}
{Abundance ratios after 13 Gyr of chemical evolution for closed box models having
 $10^4$ \msun of gas with an initial primordial metallicity, and a particular burst of
star formation resulting in 500 \msun of stars, for each of the points shown.
We display with filled triangles the results of IMF
realizations where no SNIa appeared, filled squares cases where exactly one SNIa appeared,
and filled circles those where exactly two SNIa appeared.
The thick crosses represent final abundance ratios when the IMF
is assumed to have been densely sampled, the infinite population limit.
}{fig02}

\figce{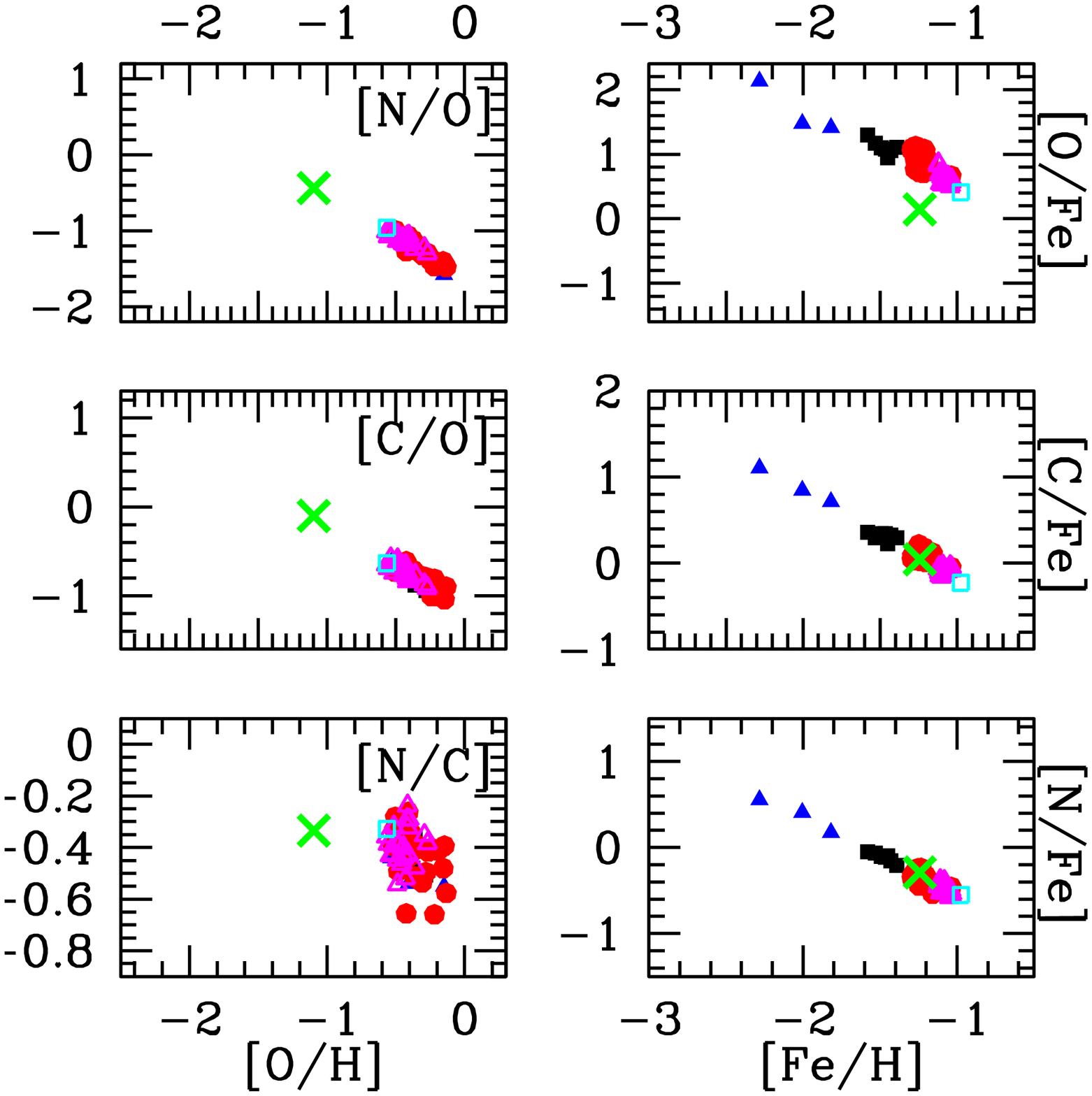}
{Abundance ratios after 13 Gyr of chemical evolution for closed box models having
 $10^{4.5}$ \msun of gas with an initial primordial metallicity, and a particular burst of
star formation resulting in 1,600 \msun of stars, for each of the points shown.
We display with filled triangles the results of IMF
realizations where no SNIa appeared, filled squares cases where exactly one SNIa appeared,
filled circles those where exactly two SNIa appeared, open triangles and open squares for the cases of
three and four, respectively.
The thick crosses represent final abundance ratios when the IMF
is assumed to have been densely sampled, the infinite population limit.
}{fig03}

We now determine the range and distribution of metallicities and metallicity ratios, associated to the non
unique and probabilistic nature of the poorly sampled IMFs which result from any given point in our
parameter space of total stellar mass and initial metallicity. We take for each 
value of the total mass being explored, and at each 
initial metallicity, a large number of distinct probabilistic IMFs, and run them through the full chemical
evolution code, following each individual star and binary for 13 Gyr, and keeping track of all
ejected materials, at all stellar phases, for both single stars and binaries.

In figure 2 we show a collection of final abundance ratios after 13 Gyr of evolution, for the gas, 
for 60 distinct statistical samplings of the same underlying probabilistic KTG IMF, all having the
same total mass of $M_{tot}=10^{4}$ \msun, all with the same initial metallicity of $Z_{i}=10^{-5}$. 
For comparison,
the result of running a standard chemical evolution model with the above parameters, under the assumption
of a densely sampled IMF, which then enters only as weighting factors on the yields, is shown by the 
cross. It is immediately apparent that very substantial intrinsic spreads of between 2 and 3 orders 
of magnitude result, as a consequence merely of the low total mass being turned into stars. The 
distribution of abundance ratios that results is not simply a normal distribution centered on 
the infinite mass limit given by the cross, but a heavily skewed scattering of points which is 
hard to describe by any smooth function. The causes of this complex distributions of points
are of course, the inherent variability of the IMF samplings that produced the stars used by
the chemical evolution model, compounded by the strong mass dependence of the yields.

To facilitate the interpretation of this and subsequent figures, we display with filled triangles the results of IMF 
realizations where no SNIa appeared, filled squares cases where exactly one SNIa appeared,
filled circles those where exactly two SNIa appeared, open triangles and open squares for the cases of
three and four, respectively. 
It is evident from the well defined
groupings of symbols seen in the right hand panels of figure 2 that the main drive behind the 
extreme resulting variability, of several orders of magnitude, is precisely the fortuitous
number of SNIa allotted to each particular realization.
If the total initial gas mass of star forming complexes in 
a small galaxy where of order $M_{tot}=10^4$ \msun, the extreme variability seen in figure 2
could be expected internally, if the initial metallicity where very low, as current models
of structure formation would suggest is the case for dSph and perhaps dIrr galaxies.

It must be noted that given the probabilistic nature
of the problem, the chemical outcome of a small star formation event can not possibly be thought
of as having a unique answer, as opposed to what happens in the infinite total mass limit. Further,
the infinite total mass limit answer is not even a good 'average' description of what is happening
at this low mass scale, not only does it not coincide with the 'most probable' outcome, indeed, 
hardly any particular realizations are found near it. 

The strong negative correlations seen in the [C/O] vs. [O/H] plot, of slope -1, is inherent to 
the quantities being plotted. In a stellar population of $ Z_{i} = 10^{-5}$,
O is produced by massive stars {with $m > 10$ \msun}
, but  C and N mainly by LIMS.
Therefore,
high final [O/H] values imply an IMF with more MS or some very massive stars with
a similar amount and type of LIMS, and therefore similar C or N final abundance.
In some models up to four MS with $m<15$ \msun are formed and in only very few models MS as 
big as 90 \msun appear.
Since C is produce also in MS but less than O, and N is mainly synthesized by LIMS
the [N/C] values show a trend to decrease slightly with increasing [O/H].
A further second order effect behind the steeper correlation in the [N/O] vs. [O/H] plot
is driven by the constraint of a fixed total stellar mass, O is produced mostly by
the massive stars, therefore having large quantities of this element implies less material 
left to make lower mass stars, and hence a somewhat reduced N production.

Together with figure 2, the following two figures, figures 3 and 4, form a sequence of increasing 
total stellar mass, at the same initial primordial metallicity of the models shown in figure 2.
Figure 3 is analogous to figure 2, but for star formation
episodes resulting in total stellar masses of 1,600 \msun. We see similar
distributions to those shown in figure 2, with the scatter diminishing, and 
slowly tending towards more regular distributions. However, although the range of expected abundance 
ratios has gone down significantly, it still covers orders of magnitude and is heavily skewed, 
the infinite population limit is not in any respect a good prediction of what one actually finds.

Figure 4 is analogous to figure 2, but for a series of star formation
bursts resulting always in a 16,000 \msun total stellar mass. As always, the underlying 
probabilistic IMF, initial primordial metallicity 
and stellar yields have been kept constant.
In this case, the distribution of resulting final metallicities for the gas in the closed box models, show
significantly less scatter than in the previous cases. Although the different results have not yet converged
to the infinite mass limit shown by the cross, and are still spread out over almost 
1 dex in [N/O], [C/O] and [O/Fe], the cloud of points is now easily 
described by simple probabilistic functions. In Cervi\~{n}o \& Luridiana (2006), 
it is shown how the distribution of integral properties of stellar populations, at fixed
total mass, resulting from the stochastic sampling of the IMF, first tends to a normal distribution, 
and then converges to the infinite population limit. This is what we see in figure 4, where
abundance ratios like [N/O] still show evidently skewed distributions, while for ratios like [C/Fe] vs. 
[Fe/H], the distributions are more symmetric about the infinite population limit.

\figce{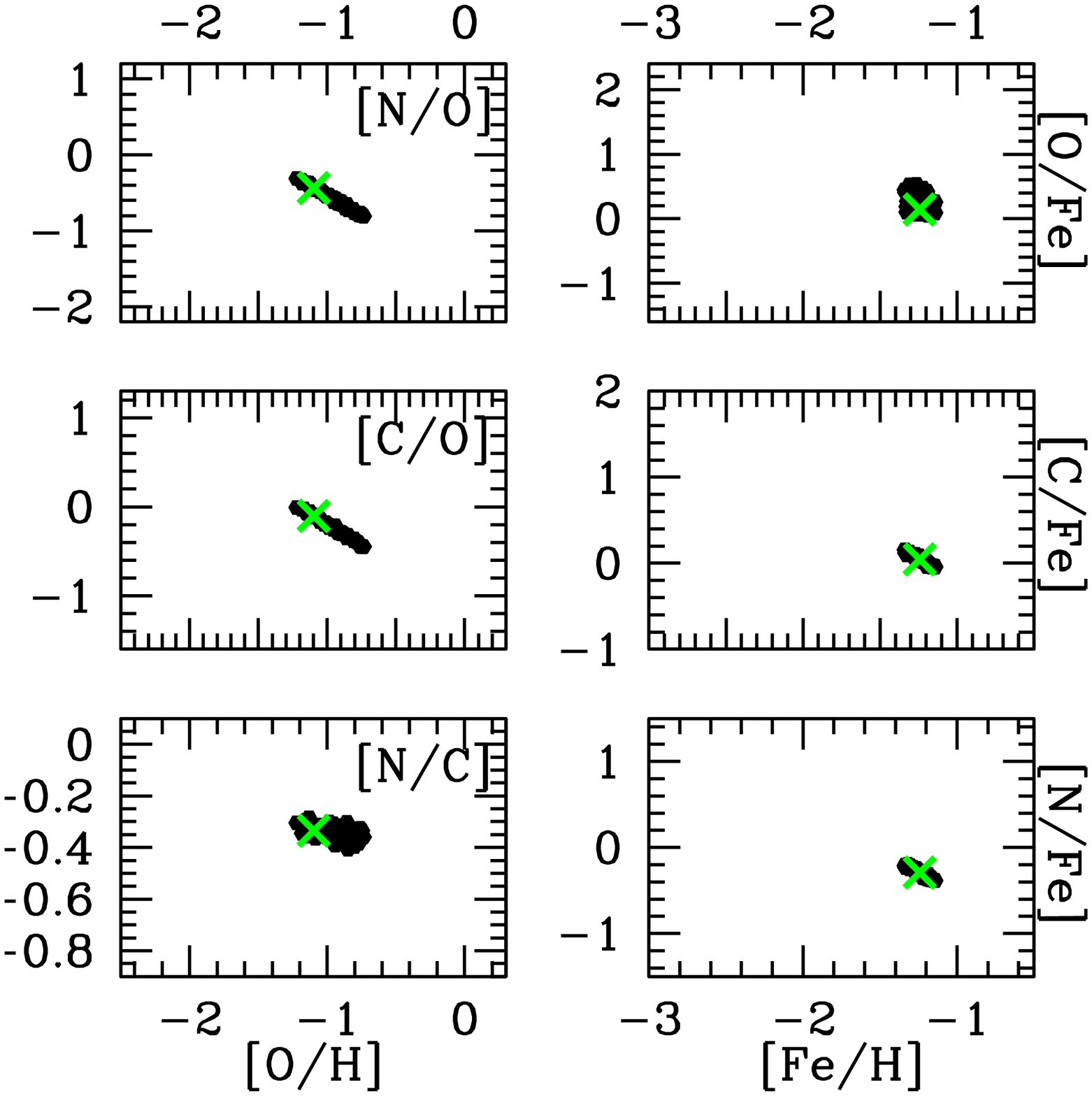}
{Abundance ratios after 13 Gyr of chemical evolution for closed box models having
$10^{5.5}$ \msun of gas with an initial primordial metallicity, and a particular burst of
star formation resulting in 16,000 \msun of stars, for each of the points shown.
The thick crosses represent final abundance ratios when the IMF
is assumed to have been densely sampled, the infinite population limit.
}{fig04}

\figce{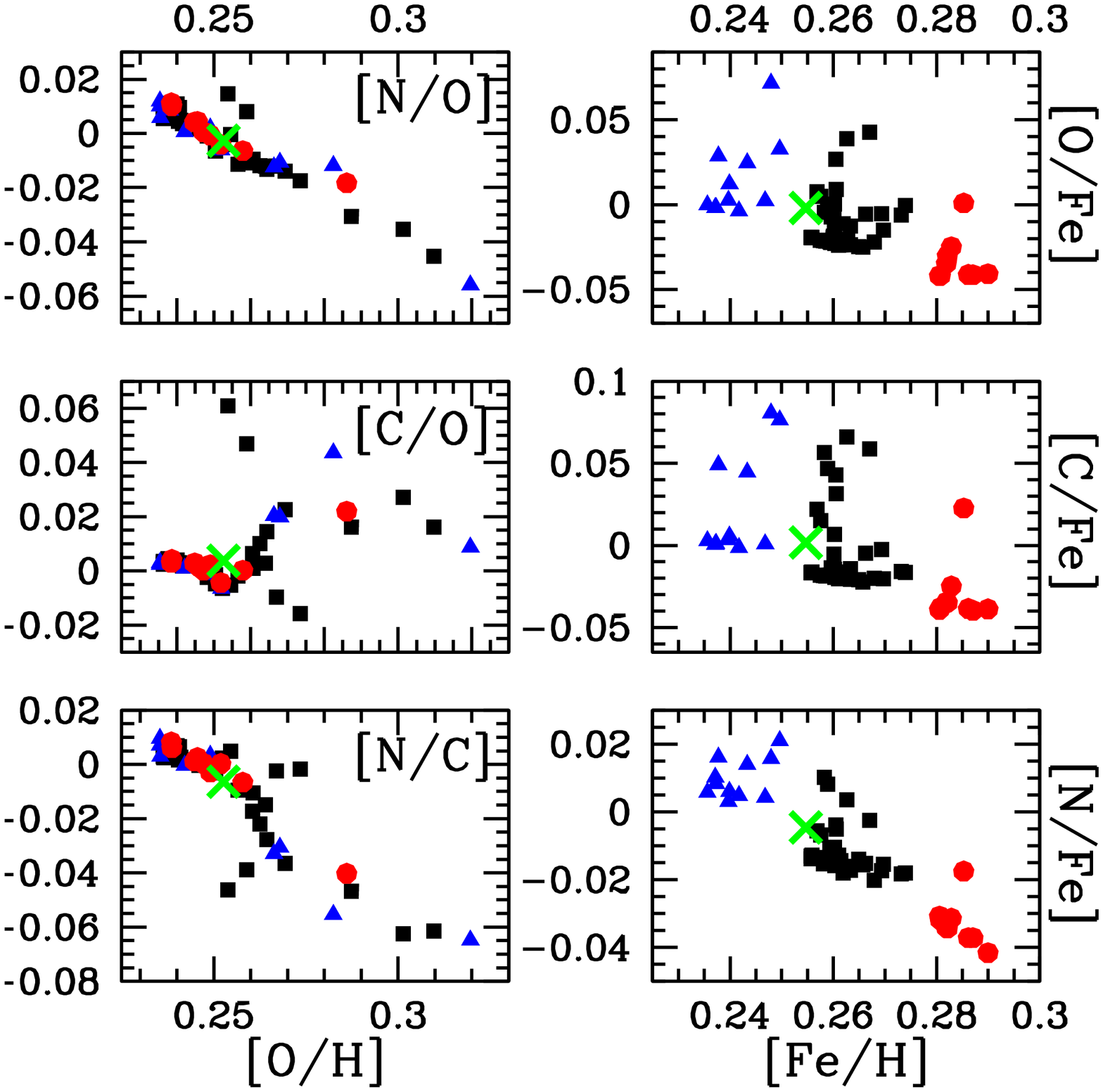}
{Abundance ratios after 13 Gyr of chemical evolution for closed box models having
 $10^4$ \msun of gas with an initial solar metallicity, and a particular burst of
star formation resulting in 500 \msun of stars, for each of the points shown.
We display with filled triangles the results of IMF
realizations where no SNIa appeared, filled squares cases where exactly one SNIa appeared,
filled circles those where exactly two SNIa appeared.
The thick crosses represent final abundance ratios when the IMF
is assumed to have been densely sampled, the infinite population limit.
}{fig05}

Figures 5 and 6 are analogous to figures 2 and 3 but show results for a series of experiments starting 
this time from solar metallicity gas. As in the previous cases, the underlying probabilistic 
IMF, initial metallicity and stellar yields have been kept constant, only the random seed
determining the stochastic sampling of the KTG IMF has been changed in going from one
particular realization to another.
In this case, the metals ejected by the stars formed are enriching gas which already has a
significant level of heavy elements present, and hence the statistical variations
in the final resulting abundances and abundance ratios, are much smaller, only of about 0.1 dex.

The correlations mentioned in connection to the previous figures are still evident, as are the groupings
of symbols showing a strong dependence of the results on the chance allotment of SNIa's for each
realization. 
Also, note that the distribution of points in figure 5 is far from smooth, and not only is it nowhere near 
converging to the infinite total mass limit given by the cross, it does not even center smoothly 
or in any easily describable manner about said point. However, in going to the larger total mass of the 
models shown in figure 6, we again see that although the distribution has not converged to the infinite
population limit, { it begins to tend towards} a much smoother situation.

We see also that as stellar lifetimes increase with increasing stellar metallicity,
in Figure 5 ($Z_{i}=0.02$),  the number of models with no SNIa's is higher than
that in figure 2 ($Z_{i}=10^{-5}$), even though 
those models were built with the same underlying IMF, and consequently the same 
probability of finding SNIa progenitors.
The correlations between [C/O] vs. [O/H], and [N/C] vs. [O/H] are different from those in figure 2,
since the MS with $Z=$ \zsun, compared to MS of primordial metallicity, eject substantially larger amounts of 
C through stellar winds.
Given the secondary origin of N, N yields for $Z_{i}=0.02$ are higher than those for $Z_{i}=10^{-5}$
but the total yield of N is particularly high for masses below 12-15 \msun, stars that eject less O.
This results in  the [N/O] vs. [O/H] correlation being less defined than for the $Z_{i}=10^{-5}$ case,
mainly for [O/H] $\sim$ 0.25-0.26.  
High final [O/H] values show cases where more MSs have enriched the ISM,
therefore leading to high [C/O] and low [N/O] and [N/C] values.

\figce{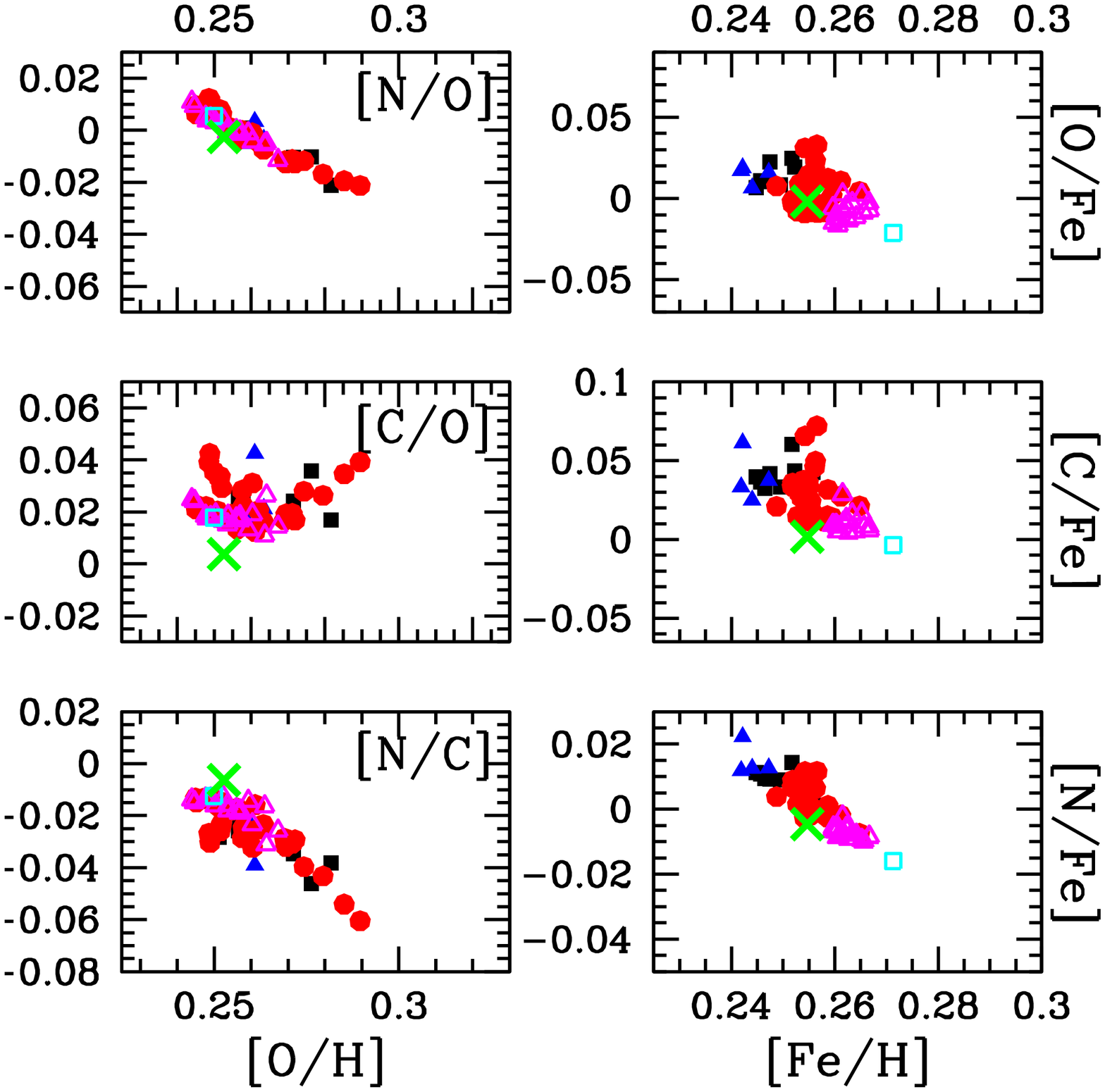}
{Abundance ratios after 13 Gyr of chemical evolution for closed box models having
$10^{4.5}$ \msun of gas with an initial solar metallicity, and a particular burst of
star formation resulting in 1,600 \msun of stars, for each of the points shown.
We display with filled triangles the results of IMF
realizations where no SNIa appeared, filled squares cases where exactly one SNIa appeared,
filled circles those where exactly two SNIa appeared, open
triangles and open squares for the cases of three and four, respectively.
The thick crosses represent final abundance ratios when the IMF
is assumed to have been densely sampled, the infinite population limit.
}{fig06}

\figce{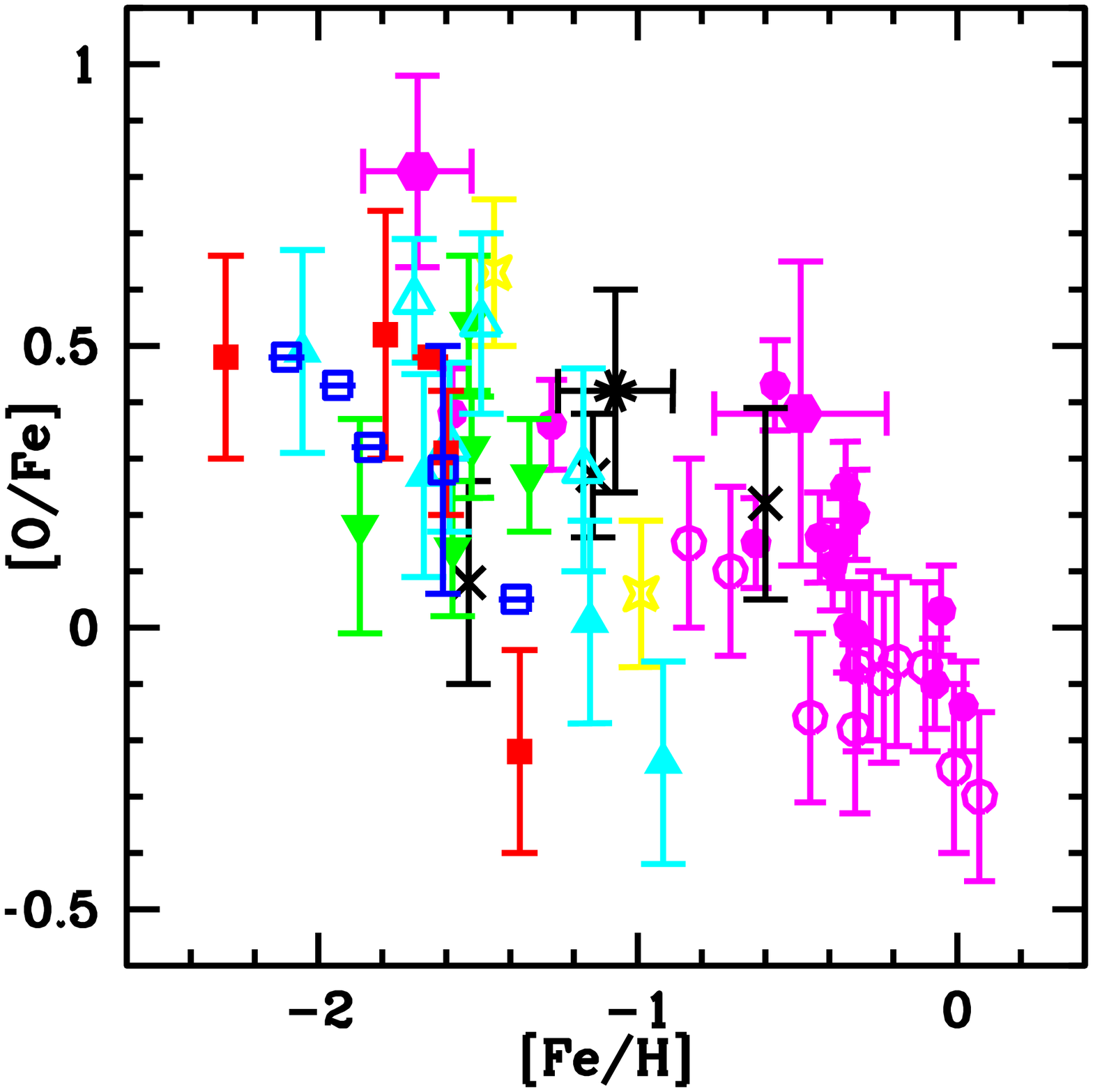}
{[O/Fe] and [Fe/H] in dSphs of the Local Group.
Abundace ratios and observational errors
from giant stars:
by Shetrone et al. (2001): {\it filled squares}: Draco, {\it open squares:}
Ursa Minor;
by Shetrone et al. (2003): {\it inverted triangles}: Carina, {\it open
triangles:} Sculptor,
{\it stars}: Leo I, {\it $\times$:} Fornax;
by McWilliam \& Smecker-Hane (2005): {\it filled circles:} Sagittarius;
by Geisler et al. (2005): {\it filled triangles:} Sculptor;
by Sbordone et al (2007): {\it open circles:} Sagittarius.
Data from planetary nebulae by Kniazev et al. (2007):
{\it big hexagones}: Sagittarius, {\it big skeletal}: Fornax.
Symbols without errors represent upper limits.
In giant stars the typical error of [Fe/H] is 0.1 dex.}
{fig07}

Also as expected,
we see that the range of values covered by the abundance ratios on the right hand side panels of 
figures 2-6, those involving Fe, is always larger than what is seen in the corresponding left hand 
side panels.
As Fe is mostly produced by SNIa's, resulting from a subset of the binaries formed and overall representing
a small fraction of the stellar population, stochastic effects are apparent here for population numbers
much larger than what is required for the distributions of metallicities produced by larger fractions
of the complete stellar population to converge. 

Resulting metallicities and abundance ratios first
display highly scattered distributions with very large dispersions, distinct groupings resulting 
from the small discrete numbers of the relevant stars responsible for their production are apparent. 
As the stellar populations increase, the distributions tend towards more regular functions,
which finally converge towards the infinite population limit. The total number of stars
at which each of the two above transitions takes place being a strong function of the fraction of
stars in the complete theoretical stellar population which contribute to the elements in 
question. Also, these transitions occur at increasingly lower total masses, as the initial 
metallicity of the gas which is being enriched by the modeled star formation burst increases, 
as seen in going from the initial primordial metallicity sequence of figures 2-4, to the solar metallicity 
sequence of figures 5-6.

\section{Observational comparisons}

We now turn to comparisons of the spread in abundance ratios shown by the models against
measured properties of local dSph galaxies.
In { figure 7} we show a compilation of the most recent [O/Fe] vs [Fe/H] 
observational data for seven dSphs of the Local Group:
Draco, Ursa Minor, Carina, Sculptor, Fornax, Leo I, and Sagittarius.
To each galaxy there corresponds a different symbol, as detailed in the figure caption.
Error bars give the reported uncertainties in the measurements of each metallicity point
for each galaxy.
We see that in this small systems, a range of metallicities is present, which can not be explained
as being due to the observational errors associated to each individual star. The dSph galaxies
summarized in this figure show unequivocal internal spreads in metallicity and abundance ratios

Most of the data were obtained from giant stars by
Shetrone et al. (2001), Shetrone et al. (2003),
McWilliam \& Smecker-Hane (2005), Geisler et al. (2005), and Sbordone et al. (2007).
McWilliam \& Smecker-Hane (2005) do not show in their figure 3
the metal-poor Sgr dSph star, I-73, ([Fe/H]=-1.03 dex) because
of its very deficient oxygen abundance ([O/Fe]=-0.91 dex),
neither do we in the left panel of our figure 7, but we consider I-73 
in the computation shown in { figure 8}.

Recently, Kniazev et al. (2007) have determined O and Fe abundances in two  
planetary nebulae of the Sagittarius dSph and one planetary nebulae in Fornax.
To account for self-pollution they corrected the [O/H] values in 2 planetary nebulae,
by 0.27 and 0.90 dex, for Fornax and the most metal poor planetary nebulae in Sgr.

Since the different data sources assume different solar abundances,
we have  homogenized the [O/Fe] and [Fe/H] values assuming
the solar abundances by Asplund et al. (2005). This correction shifts points to higher
[Fe/H] and [O/Fe] values by 0.00--0.07 dex and 0.00--0.10 dex, respectively. 

{ To better appreciate the internal dispersion in abundance ratios for each of the galaxies
in figure 7, in  figure 8} we condense each { of the galaxies shown in figure 7} to a single point, 
with the position of the symbol giving the error weighted mean of [Fe/H] and [O/Fe] for each galaxy, 
using the same symbols as in { figure 7}. The size of the bars on the symbols in { figure 8 
this time show} the error weighted standard deviation for the quantity being plotted
on each axis.  The bars in this { figure} hence give the internal spread in [Fe/H]
and [O/Fe] values found for each galaxy. In fact, these galaxies present a range of [O/Fe] values
at relatively narrow ranges in [Fe/H]. 
{ For example,} we see Sagittarius as the hexagon to the lower right, with mean values of -0.5 dex and
0.06 dex in [Fe/H] and [O/Fe], respectively, with the available collection of measurements for Sagittarius
showing an internal spread having a standard deviation of 0.9 and 0.5 in [Fe/H] and [O/Fe], respectively.
In { figure 7} we have 5 stars (4 for Ursa Minor and one for Draco) with only
upper limits in [O/Fe], these have been excluded from consideration in producing { figure 8}.
{ We see that  the usual identification of [Fe/H] as a time scale probably serves only
approximately in these cases, as we see large internal dispersions on both axes, 
with no clear causal link between them.}

Most of the dSphs show [Fe/H] $<$ -1, but Sgr is more metal rich
with  -0.9 $<$ [Fe/H] $<$ 0.
The dSphs shown present dispersions in [O/Fe] of between -0.3 and +0.8 dex, with
Draco and Sagittarius being the dSphs with the largest dispersion,
with $\sigma$ [O/Fe] $\sim$ 0.8 and 0.6 dex, respectively. 
We see the well known trend for [O/Fe] in dSph being lower than [O/Fe] values
in disk and halo stars of the solar vicinity at the same [Fe/H] values, see e.g. Venn et al. (2004).

\figce{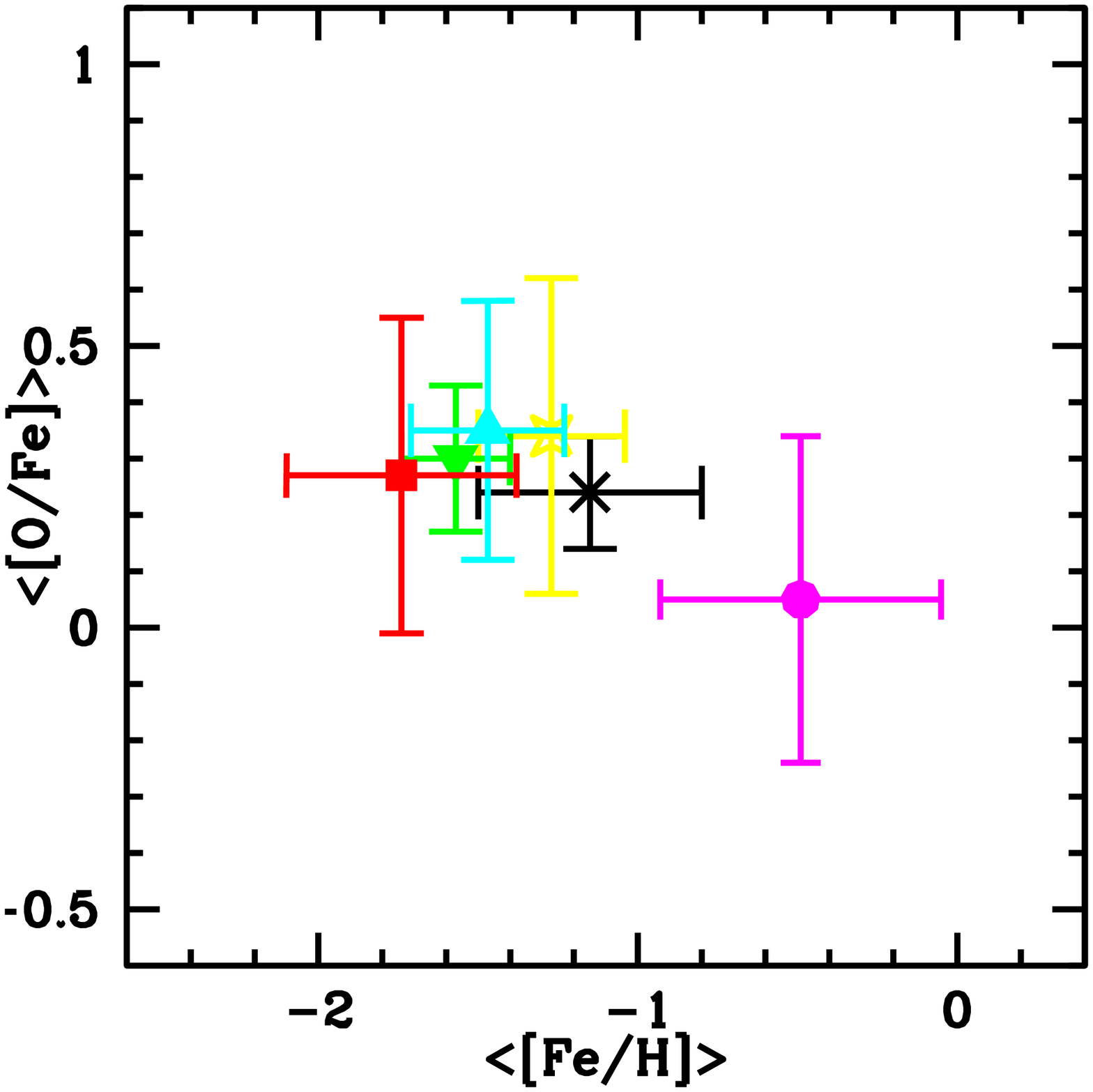}
{ Mean values and internal dispersions in [O/Fe] and [Fe/H] for the local dSph galaxies of figure 7,
considering giant stars and PN for each dSph of our sample. Symbols for each galaxy are as in figure 7.
Ursa Minor is not shown because four of its five data points are [O/Fe] upper limits.}{fig08}

We see in { figure 8} that the internal
dispersion of values reported for the local dSph's are larger than what our models
yield for primordial metallicity bursts of $M_{tot}=10^{5.5}$ \msun, but smaller than what we
obtained at $M_{tot}=10^{4.5}$ and primordial metallicity. { If one wanted to interpret all the observed internal scatter in
abundances as arising from the stochastic effects of IMF sampling, this would suggest that star formation
in dSph galaxies probably took place in star forming complexes having total masses in the range 
$10^{4.5}-10^{5.5}$ \msun.}  It is more probably the case
that other effects { which can certainly not be ruled out, and which in all probability also operate to some extent,} 
are also at work, e.g. the dynamical mixing and time evolution differences in the
expansion of SN shells of type Ia and type II explored by Marcolini et al. (2006). In any case, we see that
the effect we explored here is quite capable of resulting in significant internal abundance scatter
in dSph, as is also the case in the MW halo stars studied by Cescutti (2008). 

We also note that the average metallicities for said models agree with what 
most local dSph show, Sagittarius exempted.
We show a final set of models, analogous to those of figure 2, but this time at the intermediate
metallicity of $Z=0.004$, and a total mass of $10^{4}$ \msun, seen in figure 9. The initial metallicity
is higher than primordial, leading to smaller variations than what is seen in figure 2, but larger than
what resulted from the initial solar metallicity cases of figure 5, all at the same total mass. 
Again, in figure 9 we see the effect of the increase in stellar lifetime with increasing $Z_{i}$
on the number of SNIa enriching the ISM. The number of models with no SNIa at all
is 4, 5 and 10 for the models with $Z_{i}=10^{-5}$, $Z_{i}=0.004$, $Z_{i}=0.02$, respectively, shown by the
triangles in figures 2, 5 and 9.

\figce{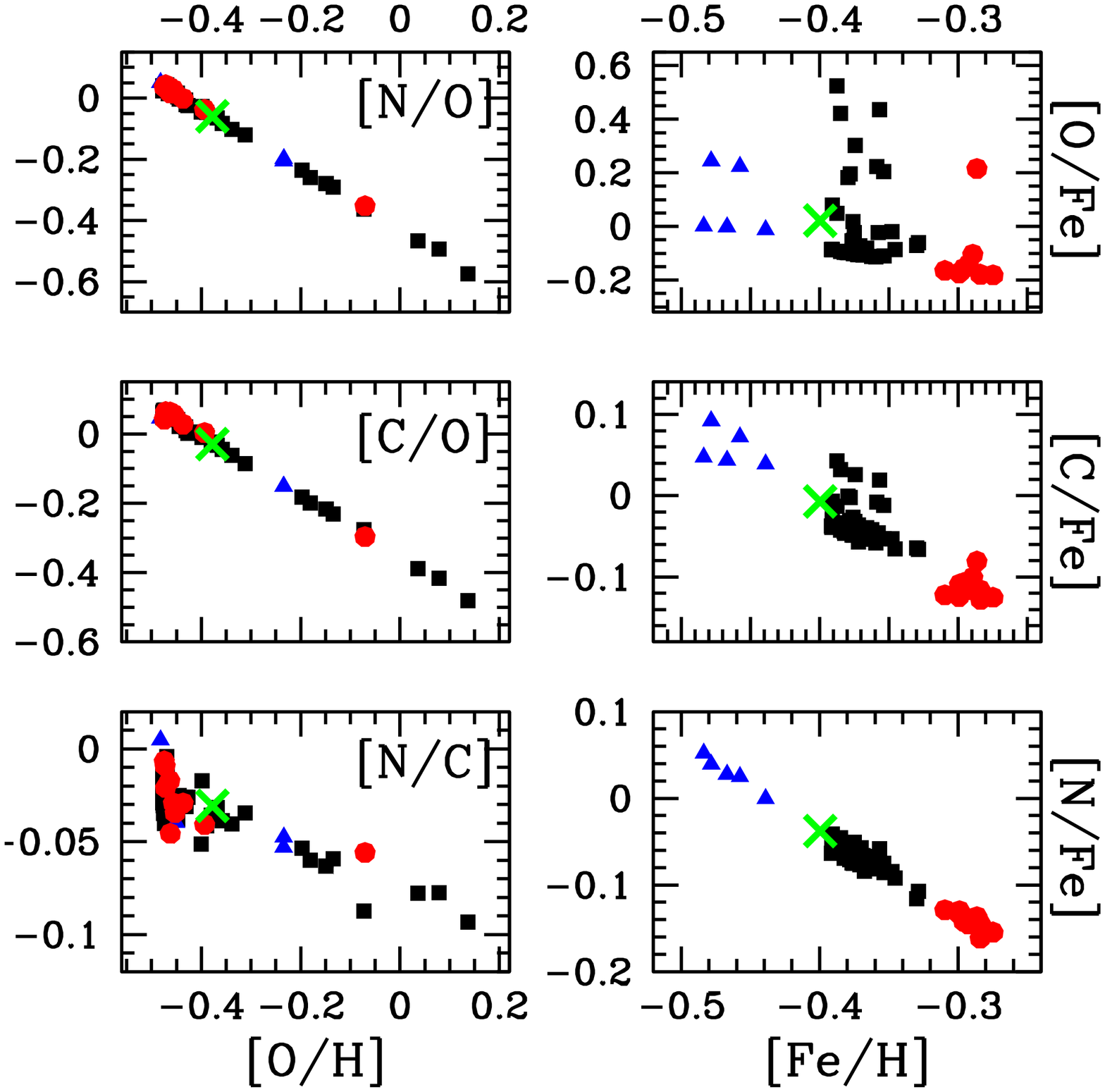}
{Abundance ratios after 13 Gyr of chemical evolution of a closed box model having
$10^{4.0}$ \msun of gas with an initial $Z=0.004$, and a particular burst of
star formation resulting in 500 \msun of stars, for each of the points shown.
We display with filled triangles the results of IMF
realizations where no SNIa appeared, filled squares cases where exactly one SNIa appeared,
and filled circles those where exactly two SNIa appeared.
The thick crosses represent final abundance ratios when the IMF
is assumed to have been densely sampled, the infinite population limit.
}{fig09}

The above case is included
to show that high internal spreads as seen in Sagittarius, at average metallicities an order of magnitude
higher than what all other dSph's in the sample show, as seen in Sagittarius, are still quite plausible
as arising from reasonable small mass star formation events. We have not attempted to tailor a particular set of
input parameters, total mass and initial metallicity, (or even to include hypothetical inflows 
or outflows of gas) to exactly reproduce the observations shown,
but it is obvious that if star formation in dSph's took place in star formation complexes
of total masses of between $3 \times 10^{4.5}$ and $3 \times 10^{5.5}$, one need not invoke changes 
in the intrinsic, underlying IMF, or any other physical effects other than the natural 
variations due to the poor sampling of the IMF, to account for the observed scatter.

A general criticism which could be made of the interpreting
the observed internal dispersion in local dSph's as arising from the intrinsic variability of a poorly
sampled IMF, is that the models we have calculated give gas metallicities, resulting from the
metal enrichment due to the stars formed, and not stellar metallicities, which is what the observations
summarized report. However, we must bear in mind that the star formation histories of local
dSph's typically display fairly complex time variations, frequently well modeled as a series
of distinct bursts. See for example, Hernandez et al. (2000), Lanfranchi et al. (2006).
Clearly, it is an initial star forming phase what we might be modeling, proceeding in small mass star formation complexes, and
enriching locally the gas which then goes on to form subsequent generations of stars, which form most
of the current stellar population in observed dSphs. That the ISM of these galaxies has been poorly mixed,
is evident from the observed dispersions in abundance ratios.

\section{Model temporal evolution}

In this last section we explore the details of the temporal evolution of some of the various
parameters which the chemical model follows through time. We begin with figure 10 which gives
the evolution of a series of abundances in the ISM in the case $M_{tot}=10^4$ \msun and an initial primordial
metallicity for the gas, the same series of models of figure 2.

The thin curves give the evolution of various [Xi/Xj] gas abundances, for the distinct 
IMF realizations, as always, at fixed underlying probabilistic IMF. 
All models start at [Xi/Xj]=0.0 when Xi and Xj are both heavy elements,
as we have scaled the initial abundances as solar, except for H and He.

The evolution of every [Xi/Xj] depends mainly on the particular collection
of massive stars formed in the stochastic IMF, and the onset of SNIa's, due to the chance allotment
of binaries in the required mass range, for every particular model.
In the panel giving [O/Fe] vs. [Fe/H] we can see that
some models present a vertical increase of [O/Fe] 
at very early times, the result of massive stars with $ m>40$ \msun, which  do not produce iron
but contribute plentiful oxygen. In other cases however, some models present a decrease in this 
quantity at early times due to massive stars with $ m<15$ \msun, which produce more iron than oxygen.
Here again, the low mass weighted IMF, and the strong leverage afforded by the high mass
weighted yields translates into large intrinsic variations in abundances and metallicities, and in the
evolution of this quantities, simply due to the statistical sampling implied by the probabilistic IMF.
For comparison, the thick curve gives the outcome for the infinite population limit, for the same
fixed initial metallicity and underlying IMF. The distribution of particular outcomes is clearly heavily
skewed, with most occurrences to one particular side of the infinite mass limit.

\figce{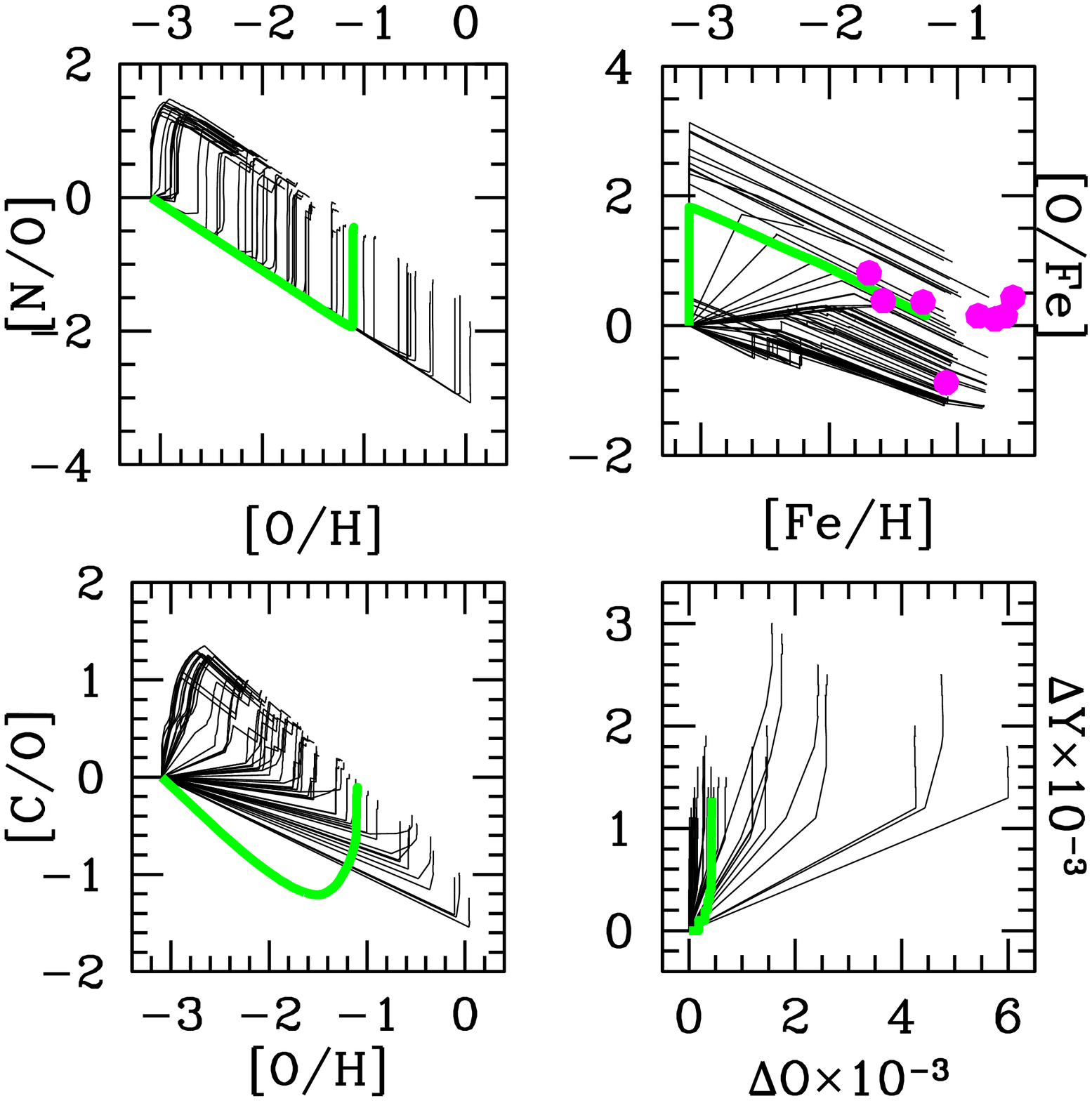}
{Evolution of abundance ratios and He vs. O values
for a series of bursts of  500 \msun enriching gas with an
initial metallicity of $10^{-5}$.
The thick lines represent the evolution at the infinite mass limit.
{ Filled circles: stars and planetary nebulae in the Sagittarius dSph for
[Fe/H]$ < -0.5$. Circle size represent the
typical error in abundance determinations, see figure 7. Star I-73 ([O/Fe]=-0.91) is included}.
}{fig10}

The general trend for [C/O] and [N/O] is similar,
because at that low $Z$, carbon and nitrogen are produced mainly in LIMS.
The delayed contribution to N and C by LIMS is shown by the increases
of [C/O] and [N/O] at high values of [O/H]. Models often present 
abrupt nearly vertical increases 
in [N/O] and [C/O] ratios, due to the explosions of individual
massive stars with { 8 $<m <$ 10 \msun}, having carbon and nitrogen yields higher
or similar to oxygen yields. 
Moreover, carbon yields for LIMS are negative for masses between 6 and 7 \msun.
The contribution of LIMS at $Z_{i}=10^{-5}$ to the [O/H] values 
becomes important when few SNII have exploded, that is the cases
for models with low [O/H] values. This explains the secondary increases in [O/H] at late times,
seen mainly in [C/O] and [N/O] for [O/H] $<$ -2.0, mainly.

\figce{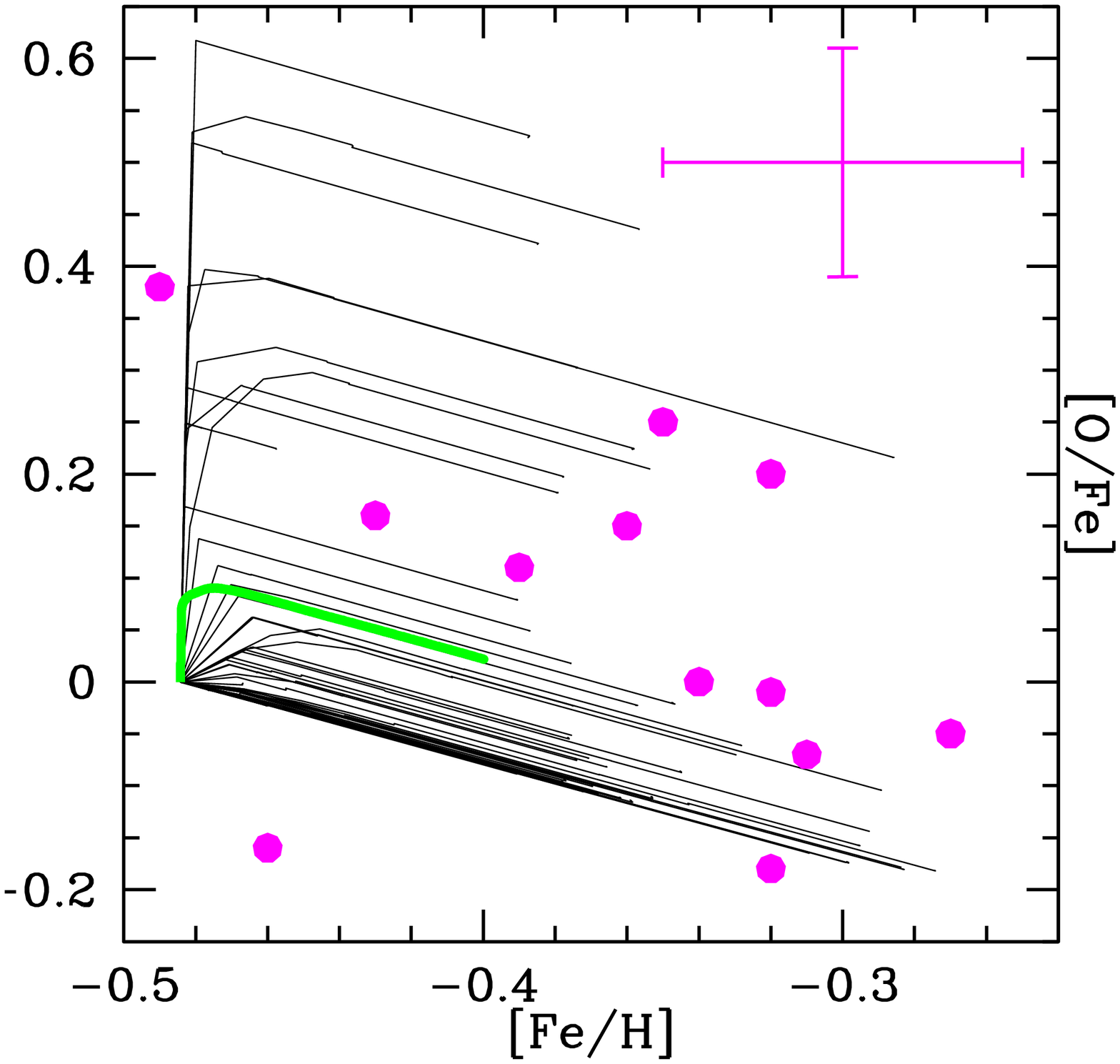}
{{ Evolution of [O/Fe] vs [Fe/H] for a series
of bursts of 500 \msun enriching gas with an initial $Z=0.004$.
The thick line represents the evolution at the infinite mass limit.
Filled circles: stars and planetary nebulae in Sagittarius dSph for
[Fe/H]$ > -0.5$. Bars represent typical error in abundance determinations,
see figure 7.}
}{fig11}

As He is produce by massive stars and LIMS, every model shows an increases in $\Delta Y$.
$\Delta O$ is higher for models { that form  massive stars or} at least one very massive star, see
Carigi \& Peimbert (2008). Again, we see a huge variation in the resulting values and
evolution of $\Delta Y$ and $\Delta O$ values, as at this low mass they are parameters
strongly sensitive to very small numbers of stars in specific critical mass ranges.

\figcw{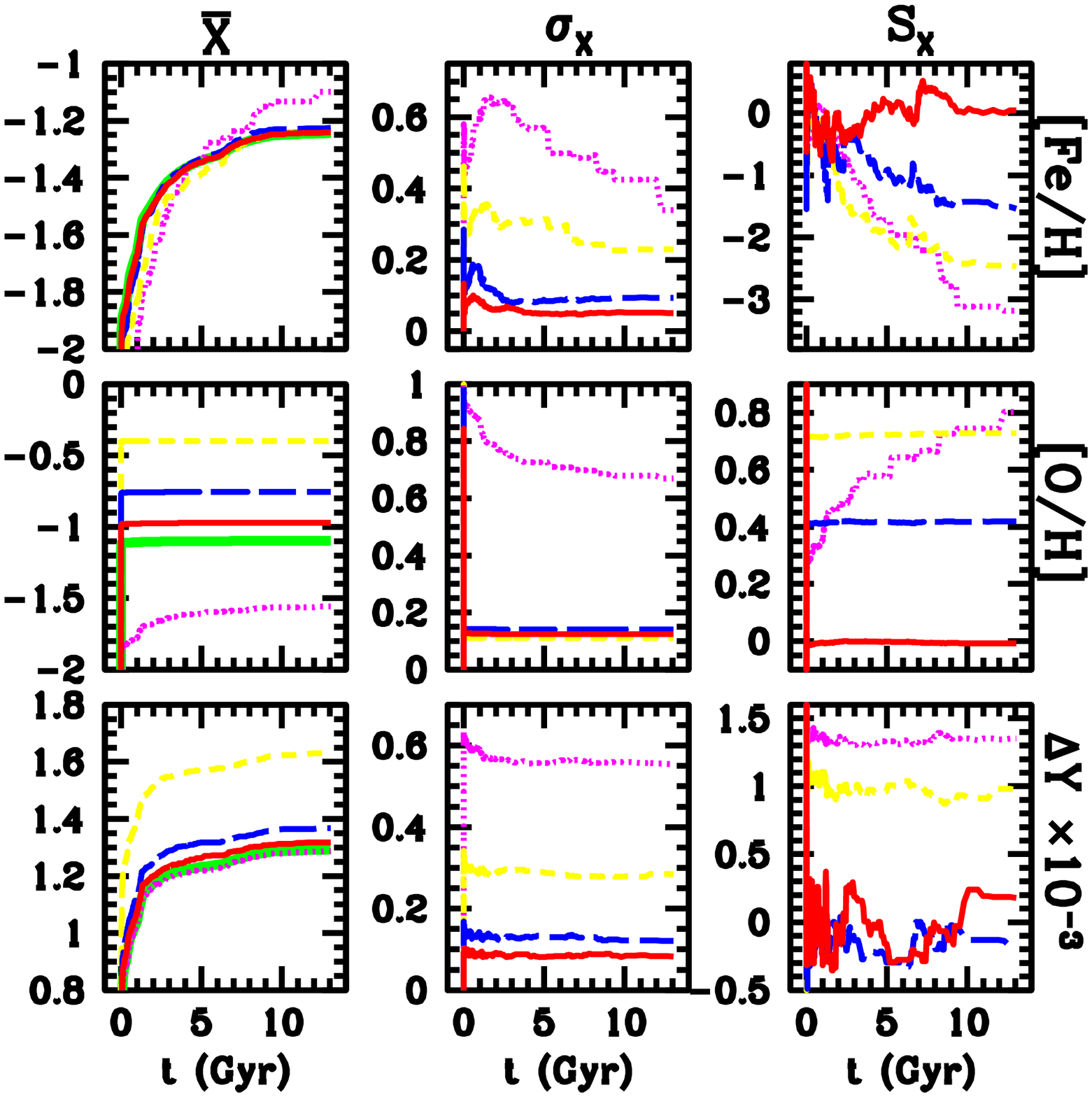}
{Evolution of mean values, standard deviation and skewness
of [Fe/H], [O/H], and $\Delta Y$ for models with initial $Z=10^{-5}$
and four different masses of the star formation burst.
Dotted lines for 500 \msun, short dashed lines for 1,600 \msun
long dashed lines for 5,000 \msun and thin continuous lines for 16,000 \msun.
In the leftmost panels the thick continuous lines represent results
at the infinite population limit
}{fig12}

{
Since some of the data collected in this paper
present [Fe/H] ratios higher than those values
shown in figure 10, we present in figure 11
 the evolution of [O/Fe] vs [Fe/H] in the
ISM for random IMF samplings of stellar populations of 500 \msun
and initial $Z=0.004$.
The general behavior is similar as that for $Z=10^{-5}$,
but the dispersion in [O/Fe] and [Fe/H] is lower for
$Z=0.004$ since the stars pollute a previously enriched gas.
}

{
The Sagittarius dSph is the galaxy with the most available data and highest
dispersion in [Fe/H] and [O/Fe] of our sample, as seen in figures 7 and 8.
Sagittarius presents at least 2  chemically separated stellar populations,
one for [Fe/H]$ < -0.5$ and another for [Fe/H] $ > -0.5$.
We have overimposed each group of stars in our predicted evolution of ISM [O/Fe] vs [Fe/H]
for $Z=10^{-5}$ and $Z=0.004$, in figures 10 and 11, respectively.
Based on the [O/Fe] vs [Fe/H] evolution shown in those figures,
we infer that the group of stars with [Fe/H]$ < -0.5$ formed from a gas enriched by stars of low $Z$,
while the second group, stars with [Fe/H]$ > -0.5$ did so in a well mixed ISM previously enriched by low $Z$ stars
and then polluted randomly by the approximately $Z=0.004$ stars of the first group.
The observed stars were probably formed in subsequent bursts at
different times and zones, with limited chemical mixing among
different galactic zones. The blurred character of the [Fe/H] axis as a time axis becomes
apparent, as distinct random realizations reach a range of final [Fe/H] values, after
being evolved for a fixed time interval.}

For the 21 sets of models we have computed the evolution
of the mean value, standard deviation and skewness defined as:

\begin{eqnarray}
S = \frac{ \sqrt {n} \sum_{i=1}^n (x_i-\bar{x})^3 }   {\left(\sum_{i=1}^n (x_i-\bar{x})^2\right)^{3/2}},
\end{eqnarray}

\noindent
for all the [Xi/Xj] elements considered and Y values.
The final two figures, figures 12 and 13, show the temporal evolution of these three important 
statistics for the ISM properties of the closed box models calculated, all for the primordial 
metallicity case, at four different total stellar masses, 500, 1,600, 5,000 and 16,000 \msun,
given by the dotted, short dashed, long dashed and continuous curves, respectively.
Again, for comparison the infinite population limit is given by the thick continuous curves, 
only applicable to the leftmost panels where the mean values are displayed.

In both figures we see a clear trend in going from the small to the large total masses, 
the mean values tend to those of the infinite population limits, with both the standard deviations 
and skewness falling towards zero, as expected. That the means appear away from the infinite population limit 
is due to the fact that we have only simulated a limited
and relatively small number ($\sim 60$) of bursts at each mass, the trends seen in the standard deviations
and skewness however, are rather robust to the total number of models run at each mass. At the smallest
total stellar masses of 500 and 1,600 \msun, we see large dispersions, and very large values for the skewness,
as inspection of the previous figures would have suggested.

For the lower two values of the total stellar mass,
the standard deviation decreases as a function of time,
mainly for [O/H] and [Fe/H],
due to the delayed contribution of LIMS and SNIa, as
the stochastic IMFs are mostly filled with low mass stars.
That is, the stochastic effects are more important for massive stars
and therefore at the beginning of the evolution.
The  evolution of the standard deviation of [O/H] and [Fe/H] determines
the evolution of the standard deviation of the other abundances ratios
shown in figure 13.

The skewness for [O/H] and $\Delta Y$ is positive
because O and He are produce by MS and
both these yields increase with the stellar mass for $Z=10^{-5}$.
Since the stochastic IMF is weighted towards low stellar masses, even more in the case of
$M_{tot}=10^4$ \msun, there are few models with very massive stars and therefore
high [O/H] and He values. That produces a positive tail in the distribution
for [O/H] and $\Delta Y$.
The positive skewness in O/H produces the negative skewness of [N/O] and [C/O].

We see in figure 12 that the skewness for [Fe/H] is negative, this is due to the nature of the
distribution function of 
secondary masses for the binary system progenitors of SNIa, $f(m2)$.
This $f(m2)$ is heavily weighted towards intermediate mass secondary stars, 
rather than to low mass secondary stars.
Therefore, in our models most SNIa come from secondary stars with short lifetimes.
Since the lifetime of these SNIa progenitors is determined by the lifetime of the $m2$ star, 
there are, especially at early times, many more models with many SNIa than models with no SNIa.
This implies that most models result in high [Fe/H] values and 
only few models in low [Fe/H] values, those with zero or one SNIa.
This last produces a negative tail in the [Fe/H] distribution.
This tail becomes more negative along the evolution (absolute values 
for the skewness of [Fe/H] increasing with time)
because new SNIa explode with the passing of time in some models,
reducing the number of models with no SNIa, the case of 500 \msun total stellar mass, 
or few SNIa, in the cases of 1,600 \msun and 5,000 \msun total stellar masses.
Consequently, the skewness for the [O/Fe] distribution is positive, due to the skewness in the [O/H]
distribution being positive and that in the [Fe/H] distribution negative.

The skewness increase with time for [O/H] in the models with a total stellar mass of 500 \msun,
due to the O contribution of LIMS for $t >$ 1 Gyr.
O is mainly produced by MS, when the stellar population has low total mass,
the number of SNII is low and that causes low [O/H] values.
For populations with $Zi=10^{-5}$ the LIMS contribution to O enrichment is  delayed. 
Since the O yields of LIMS are lower 
than the O yields of MS, by orders of magnitude,
LIMS can contribute appreciably to [O/H] values only in the cases when few
SNII appear. Since all models have similar numbers and types of LIMS, but
only few models have very massive stars (high [O/H] values) the
positive tail of the distribution increase with time.

\figcw{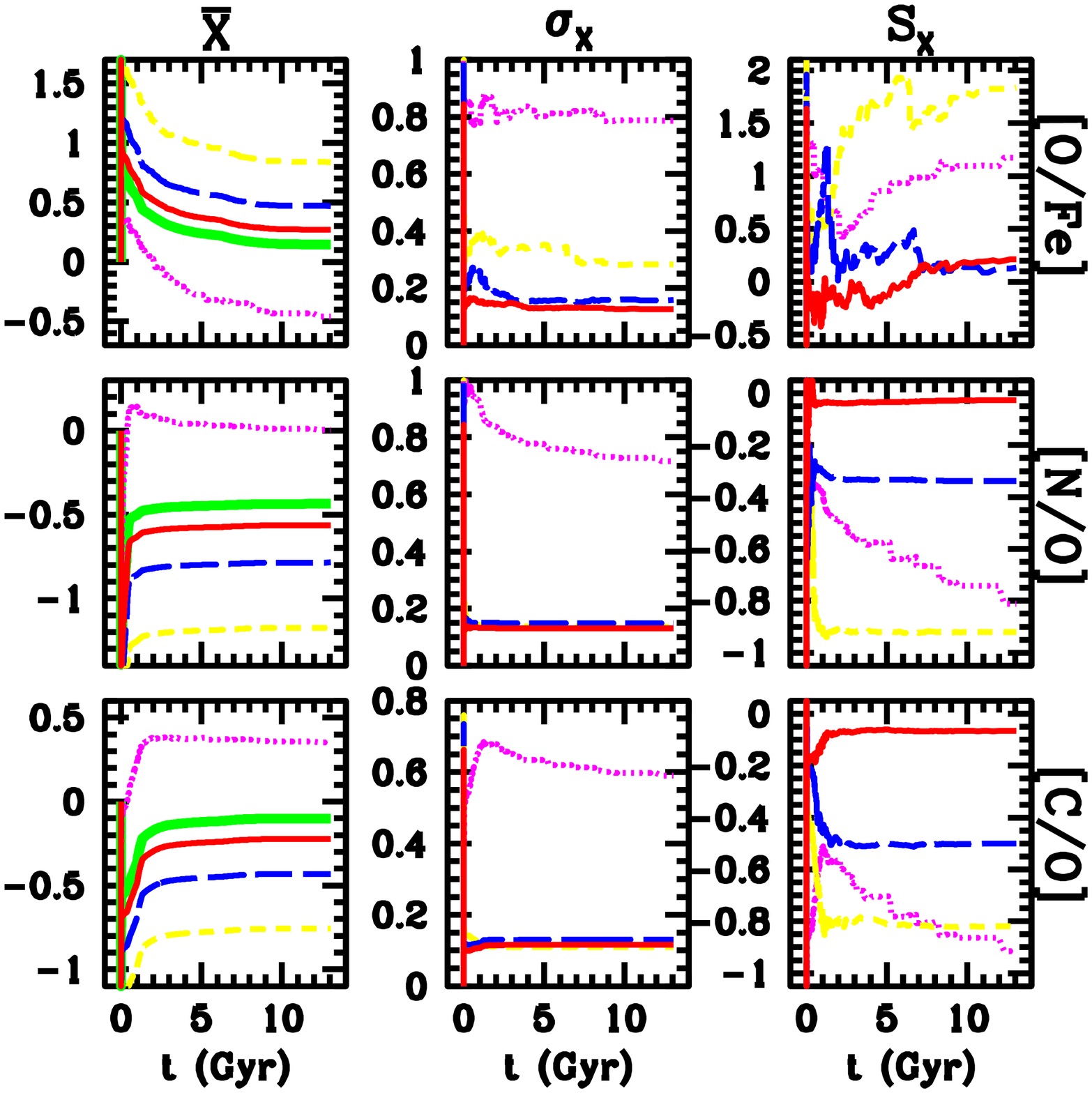}
{Evolution of mean values, standard deviation and skewness
of [O/Fe], [N/O], and [C/O] for models with initial $Z=10^{-5}$
at four different masses of the star formation burst.
Dotted lines for 500 \msun, short dashed lines for 1,600 \msun,
long dashed lines for 5,000 \msun and thin continuous lines for 16,000 \msun.
In the leftmost panels the thick continuous lines represent results
at the infinite population limit.
}{fig13}

\section{Conclusions}
After performing discrete samplings of a fixed IMF, and tracing the chemical
consequences of large samples of resulting stellar populations, we conclude the
following:

For a fixed underlying IMF, a wide variety of metallicities and abundance ratios
can result from star formation events of fixed total mass.

The distributions of such metallicities and abundance ratios goes from discrete 
groupings of points covering 2-3 orders of magnitude, to heavily skewed distributions
ranging over about 1 order of magnitude, to approximately normal distributions
centered on the infinite population limit, as the total masses involved are increased.

The regimes where the above transitions take place, as the details of the  resulting distributions,
are heavily dependent on the initial
metallicity of the gas being considered, with stochastic effects growing as the initial
metallicity decreases. 

As expected, the stochastic effects are felt differently for different heavy elements, 
those most affected being the ones which result from smaller fractions of stars.

The temporal evolution of the parameters studied in the ISM generally shows a slight tendency 
towards the infinite mass population limit as time increases, a result of increasing fractions
of the stellar population reaching their relevant lifetimes.

{ It appears reasonable to infer that stochastic effects inherent to the probabilistic nature
of the IMF, becoming conspicuous at the low mass star formation, low initial metalicity regimes
relevant to local dSph galaxies, might contribute to the observed dispersion in 
abundance ratios seen in these systems.}

\section*{Acknowledgments}
{ The authors acknowledge constructive criticism from an anonymous referee
which helped in reaching a clear and balanced presentation.} 
LC´s work is partly supported by the CONACyT { grants 46904 and 60354}. 
The work of XH is partly supported by UNAM-DGAPA grant IN 114107.
The authors acknowledge helpful discussions with David Valls-Gabaud, Miguel
Cervi\~{n}o, and Manuel Peimbert on the subjects treated in this paper.

\label{lastpage}

\end{document}